\documentclass[TOTEM]{cernphprep}

\usepackage{fancyhdr, graphicx}
\usepackage{float}
\usepackage{color}
\usepackage{amsmath,amssymb,wasysym}
\usepackage{multirow}
\usepackage{epstopdf}
\usepackage{lineno}
\usepackage[numbers,sort&compress]{natbib}

\DeclareMathAlphabet{\mathcal}{OMS}{cmsy}{m}{n}

\setbox123\hbox{\small$0$}
\def\S{\hbox to\wd123{\hss}}
\setbox124\hbox{\small$_{0}$}
\def\s{\hbox to\wd124{\hss}}

\def\etal{et al.}

\def\Instline#1#2{%
	\expandafter\write1{\string\newlabel{#1}{{#1}{}}}%
	\hbox to\hsize{\strut\hss$^{#1}$#2\hss}
}

\begin{document}

\begin{titlepage}

\renewcommand{\EXPLOGO}{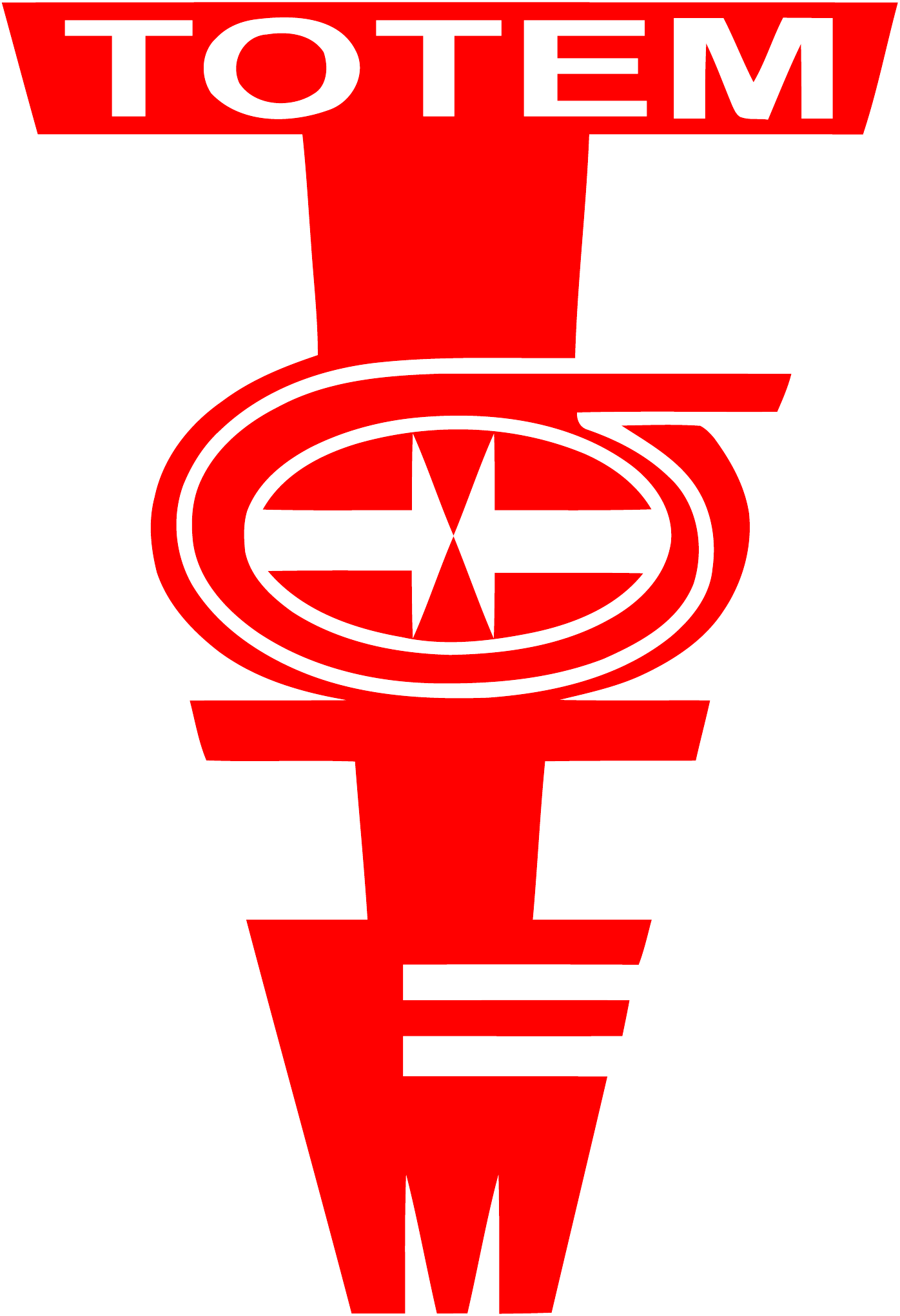}

\PHnumber{CERN-EP-2018-338}
\PHdate{\today}

\EXPnumber{TOTEM-2018-003}
\EXPdate{\today}

\title{Elastic differential cross-section measurement at $\sqrt{s}=13$~TeV by TOTEM}
\ShortTitle{Elastic differential cross-section measurement at $\sqrt{s}=13$~TeV by TOTEM}

\Collaboration{The TOTEM Collaboration}
\ShortAuthor{The TOTEM Collaboration (G.~Antchev \emph{\etal})}

\def\DeclareAuthors{%
	\AddAuthor{G.~Antchev}{}{}{a}%
	\AddAuthor{P.~Aspell}{9}{}{}%
	\AddAuthor{I.~Atanassov}{}{}{a}%
	\AddAuthor{V.~Avati}{7}{}{}%
	\AddAuthor{J.~Baechler}{9}{}{}%
	\AddAuthor{C.~Baldenegro~Barrera}{11}{}{}%
	\AddAuthor{V.~Berardi}{4a}{4b}{}%
	\AddAuthor{M.~Berretti}{2a}{}{}%
	\AddAuthor{E.~Bossini}{6b}{}{}%
	\AddAuthor{U.~Bottigli}{6b}{}{}%
	\AddAuthor{M.~Bozzo}{5a}{5b}{}%
	\AddAuthor{H.~Burkhardt}{9}{}{}%
	\AddAuthor{F.~S.~Cafagna}{4a}{}{}%
	\AddAuthor{M.~G.~Catanesi}{4a}{}{}%
	\AddAuthor{M.~Csan\'{a}d}{3a}{}{b}%
	\AddAuthor{T.~Cs\"{o}rg\H{o}}{3a}{3b}{}%
	\AddAuthor{M.~Deile}{9}{}{}%
	\AddAuthor{F.~De~Leonardis}{4c}{4a}{}%
	\AddAuthor{M.~Doubek}{1c}{}{}%
	\AddAuthor{D.~Druzhkin}{9}{}{}%
	\AddAuthor{K.~Eggert}{10}{}{}%
	\AddAuthor{V.~Eremin}{}{}{c}%
	\AddAuthor{F.~Ferro}{5a}{}{}%
	\AddAuthor{A.~Fiergolski}{9}{}{}%
	\AddAuthor{F.~Garcia}{2a}{}{}%
	\AddAuthor{V.~Georgiev}{1a}{}{}%
	\AddAuthor{S.~Giani}{9}{}{}%
	\AddAuthor{L.~Grzanka}{7}{}{}%
	\AddAuthor{J.~Hammerbauer}{1a}{}{}%
	\AddAuthor{V.~Ivanchenko}{8}{}{}%
	\AddAuthor{J.~Ka\v{s}par}{6a}{1b}{}%
	\AddAuthor{J.~Kopal}{9}{}{}%
	\AddAuthor{V.~Kundr\'{a}t}{1b}{}{}%
	\AddAuthor{S.~Lami}{6a}{}{}%
	\AddAuthor{G.~Latino}{6b}{}{}%
	\AddAuthor{R.~Lauhakangas}{2a}{}{}%
	\AddAuthor{C.~Lindsey}{11}{}{}%
	\AddAuthor{M.~V.~Lokaj\'{\i}\v{c}ek}{1b}{}{}%
	\AddAuthor{L.~Losurdo}{6b}{}{}%
	\AddAuthor{M.~Lo~Vetere}{5b}{5a}{+}%
	\AddAuthor{F.~Lucas~Rodr\'{i}guez}{9}{}{}%
	\AddAuthor{M.~Macr\'{\i}}{5a}{}{}%
	\AddAuthor{M.~Malawski}{7}{}{}%
	\AddAuthor{N.~Minafra}{11}{}{}%
	\AddAuthor{S.~Minutoli}{5a}{}{}%
	\AddAuthor{T.~Naaranoja}{2a}{2b}{}%
	\AddAuthor{F.~Nemes}{9}{3a}{}%
	\AddAuthor{H.~Niewiadomski}{10}{}{}%
	\AddAuthor{T.~Nov\'{a}k}{3b}{}{}%
	\AddAuthor{E.~Oliveri}{9}{}{}%
	\AddAuthor{F.~Oljemark}{2a}{2b}{}%
	\AddAuthor{M.~Oriunno}{}{}{d}%
	\AddAuthor{K.~\"{O}sterberg}{2a}{2b}{}%
	\AddAuthor{P.~Palazzi}{9}{}{}%
	\AddAuthor{V.~Passaro}{4c}{4a}{}%
	\AddAuthor{Z.~Peroutka}{1a}{}{}%
	\AddAuthor{J.~Proch\'{a}zka}{1c}{}{}%
	\AddAuthor{M.~Quinto}{4a}{4b}{}%
	\AddAuthor{E.~Radermacher}{9}{}{}%
	\AddAuthor{E.~Radicioni}{4a}{}{}%
	\AddAuthor{F.~Ravotti}{9}{}{}%
	\AddAuthor{E.~Robutti}{5a}{}{}%
	\AddAuthor{C.~Royon}{11}{}{}%
	\AddAuthor{G.~Ruggiero}{9}{}{}%
	\AddAuthor{H.~Saarikko}{2a}{2b}{}%
	\AddAuthor{A.~Scribano}{6a}{}{}%
	\AddAuthor{J.~Smajek}{9}{}{}%
	\AddAuthor{W.~Snoeys}{9}{}{}%
	\AddAuthor{J.~Sziklai}{3a}{}{}%
	\AddAuthor{C.~Taylor}{10}{}{}%
	\AddAuthor{E.~Tcherniaev}{8}{}{}%
	\AddAuthor{N.~Turini}{6b}{}{}%
	\AddAuthor{V.~Vacek}{1c}{}{}%
	\AddAuthor{J.~Welti}{2a}{2b}{}%
	\AddAuthor{J.~Williams}{11}{}{}%
	\AddAuthor{R.~Ciesielski}{12}{}{}%
}


\def\DeclareInstitutes{%
	\AddInstitute{1a}{University of West Bohemia, Pilsen, Czech Republic.}
	\AddInstitute{1b}{Institute of Physics of the Academy of Sciences of the Czech Republic, Prague, Czech Republic.}
	\AddInstitute{1c}{Czech Technical University, Prague, Czech Republic.}
	\AddInstitute{2a}{Helsinki Institute of Physics, University of Helsinki, Helsinki, Finland.}
	\AddInstitute{2b}{Department of Physics, University of Helsinki, Helsinki, Finland.}
	\AddInstitute{3a}{Wigner Research Centre for Physics, RMKI, Budapest, Hungary.}
	\AddInstitute{3b}{EKU KRC, Gy\"ongy\"os, Hungary.}
	\AddInstitute{4a}{INFN Sezione di Bari, Bari, Italy.}
	\AddInstitute{4b}{Dipartimento Interateneo di Fisica di Bari, Bari, Italy.}
	\AddInstitute{4c}{Dipartimento di Ingegneria Elettrica e dell'Informazione - Politecnico di Bari, Bari, Italy.}
	\AddInstitute{5a}{INFN Sezione di Genova, Genova, Italy.}
	\AddInstitute{5b}{Universit\`{a} degli Studi di Genova, Italy.}
	\AddInstitute{6a}{INFN Sezione di Pisa, Pisa, Italy.}
	\AddInstitute{6b}{Universit\`{a} degli Studi di Siena and Gruppo Collegato INFN di Siena, Siena, Italy.}
	\AddInstitute{7}{AGH University of Science and Technology, Krakow, Poland.}
	\AddInstitute{8}{Tomsk State University, Tomsk, Russia.}
	\AddInstitute{9}{CERN, Geneva, Switzerland.}
	\AddInstitute{10}{Case Western Reserve University, Dept.~of Physics, Cleveland, OH, USA.}
	\AddInstitute{11}{The University of Kansas, Lawrence, USA.}
	\AddInstitute{12}{Rockefeller Univeristy, New York, USA.}
}

	
\def\DeclareExternalInstitutes{%
	\AddExternalInstitute{a}{INRNE-BAS, Institute for Nuclear Research and Nuclear Energy, Bulgarian Academy of Sciences, Sofia, Bulgaria.}
	\AddExternalInstitute{b}{Department of Atomic Physics, ELTE University, Budapest, Hungary.}
	\AddExternalInstitute{c}{Ioffe Physical - Technical Institute of Russian Academy of Sciences, St.~Petersburg, Russian Federation.}
	\AddExternalInstitute{d}{SLAC National Accelerator Laboratory, Stanford CA, USA.}
	\AddExternalInstitute{+}{Deceased.}
}



\newif\ifFirstAuthor
\FirstAuthortrue

\def\AddAuthor#1#2#3#4{%
	\def\PriAf{#2}%
	\def\SecAf{#3}%
	\def\ExtAf{#4}%
	\def\empty{}%
	\ifFirstAuthor
		\FirstAuthorfalse
	\else
		,
	\fi
	\ifx\PriAf\empty
		#1\Aref{#4}%
	\else
		\ifx\SecAf\empty
			\ifx\ExtAf\empty
				#1\Iref{#2}%
			\else
				#1\IAref{#2}{#4}%
			\fi
		\else
			\ifx\ExtAf\empty
				#1\IIref{#2}{#3}%
			\else
				#1\IIAref{#2}{#3}{#4}%
				\relax
			\fi
		\fi
	\fi
}


\def\AddCorrespondingAuthor#1#2#3#4#5#6{%
	\AddAuthor{#1}{#2}{#3}{*}%
	\Anotfoot{*}{#5 E-mail address: #6.}
}


\def\AddInstitute#1#2{%
	\expandafter\write1{\string\newlabel{#1}{{#1}{}}}%
	\hbox to\hsize{\strut\hss$^{#1}$#2\hss}%
}


\def\AddExternalInstitute#1#2{%
	\Anotfoot{#1}{#2}%
}



\begin{Authlist}
	\DeclareAuthors
\end{Authlist}

\DeclareInstitutes
\hbox to\hsize{\strut\hss} 
\DeclareExternalInstitutes

\author{\null}
\date{\today}
\begin{abstract}
The TOTEM collaboration has measured the elastic proton-proton differential cross section ${\rm d}\sigma/{\rm d}t$ at $\sqrt{s}=13$~TeV LHC energy using dedicated $\beta^{*}=90$~m beam optics. The Roman Pot detectors
were inserted to 10$\sigma$ distance from the LHC beam, which allowed the measurement of the range $[0.04$~GeV$^{2}$$; 4$~GeV$^{2}$$]$ in four-momentum transfer squared $|t|$. The efficient data acquisition allowed to collect about 10$^{9}$ elastic
events to precisely measure the differential cross-section including the diffractive minimum (dip), the subsequent maximum (bump) and the large-$|t|$ tail. The average nuclear slope has been found to be $B=(20.40 \pm 0.002^{\rm stat} \pm 0.01^{\rm syst})~$GeV$^{-2}$ in the $|t|$-range
$0.04~$GeV$^{2}$ to $0.2~$GeV$^{2}$. The dip position is $|t_{\rm dip}|=(0.47 \pm 0.004^{\rm stat} \pm 0.01^{\rm syst})~$GeV$^{2}$. The differential cross section ratio at the bump vs. at the dip $R=1.77\pm0.01^{\rm stat}$ has been measured with high precision. The series of TOTEM elastic pp measurements show that the dip is a permanent feature of the pp differential cross-section at the TeV scale.

\end{abstract}

\end{titlepage}

\maketitle


\section{Introduction}

This paper presents a high-statistics proton-proton elastic differential cross-section ${\rm d}\sigma/{\rm d}t$ measurement by the TOTEM experiment at a center-of-mass LHC energy
$\sqrt{s} = 13$~TeV. The square of four-momentum transferred in the elastic process, $|t|$, covers an unprecedented range from $0.04$~GeV$^{2}$ to $4$~GeV$^{2}$. The
large $|t|$-spectrum has been achieved with special and efficient data acquisition, which allowed to collect an order of 10$^{9}$ elastic events. The elastic differential cross-section ${\rm d}\sigma/{\rm d}t$
spans ten orders of magnitude in one data set, providing a unique insight into the elastic interaction of protons. 

The TOTEM collaboration has already measured proton-proton elastic scattering at several LHC energies: $\sqrt{s}=2.76$~TeV, 7~TeV, 8~TeV and 13 TeV~\cite{Antchev:2013paa,Antchev:2016vpy,Antchev:2011vs,Antchev:2013iaa,DIS2017_proceedings,Paper_2p76}.
The present results continue the series of measurements at $\sqrt{s} = 13$~TeV, showing the exponential-like part at low-$|t|$, characterized by an average nuclear slope $B$, the diffractive minimum of the  ${\rm d}\sigma/{\rm d}t$ and
the perturbative regime.

The main features of the observed ${\rm d}\sigma/{\rm d}t$ at the Intersecting Storage Ring (ISR) about 40 years ago are all present at the TeV scale~\cite{Amaldi:1979kd}. The $\sqrt{s}$ dependence of the ${\rm d}\sigma/{\rm d}t$ shows
the shrinkage of the elastic peak with increasing $\sqrt{s}$, thus the average nuclear slope $B$ increases and the dip moves to lower $|t|$ values. The precise data at $\sqrt{s} = 13$~TeV confirms the significant deviation from
an exponential in the $|t|$-range from about $0.05$~GeV$^{2}$ to $0.2$~GeV$^{2}$, first observed at 8~TeV by the TOTEM experiment~\cite{Antchev:2015zza}. 

The TOTEM measurements confirmed the existence of the dip at the collision energies $\sqrt{s}=2.76$~TeV, 7~TeV, 8~TeV and 13 TeV. In total a range of $10$~TeV center-of-mass energy is covered, and the observations
demonstrate that the diffractive minimum is a permanent structure at the TeV scale~\cite{Antchev:2013paa,Antchev:2016vpy,Antchev:2011vs,Antchev:2013iaa,DIS2017_proceedings,Paper_2p76}.

\section{Experimental setup}

\begin{figure}[H]
		\centering
		\includegraphics[trim = 0mm 510mm 610mm 0mm, clip, width=1.0\columnwidth]{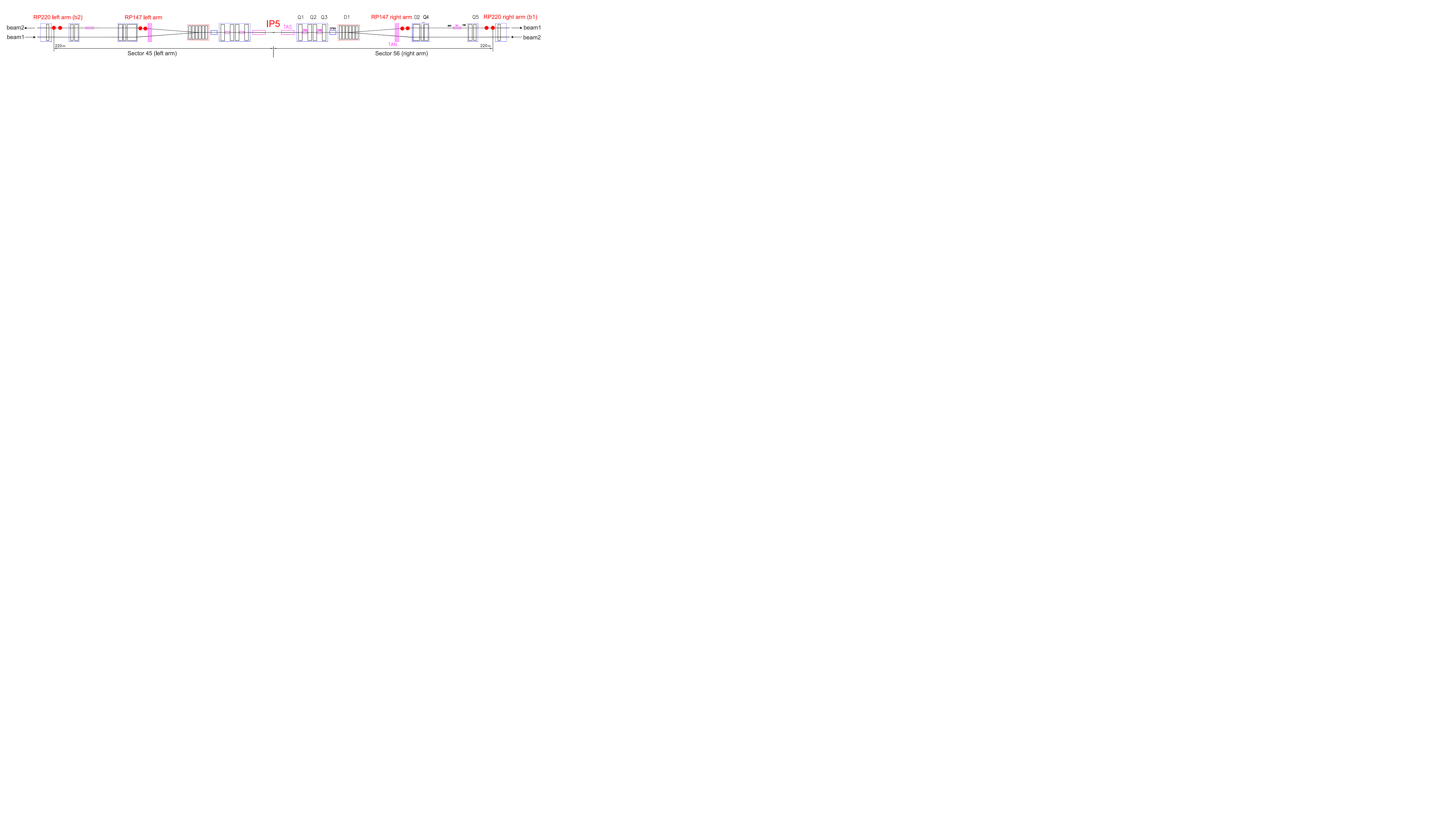}
		\caption{(color) Schematic layout of the LHC from IP5 up to the “near” and “far” Roman Pot units, where the near and far pots are indicated by full (red) dots on beams 1 and 2.}
		\label{layout_LHC}
 	\end{figure}

			The Roman Pot (RP) units used for the present measurement are located on both sides of the LHC Interaction Point 5 (IP5) at distances of $\pm213$~m (near) and $\pm220$~m (far), see Fig.~\ref{layout_LHC}. A unit consists of 3 RPs, two approaching the outgoing beam vertically and one horizontally.
			The horizontal RP overlaps with the two verticals and allows for a precise relative alignment of the detectors within the unit. The $7$~m long lever arm between the near and the far RP units has the important advantage that the local track angles in the $x$ and $y$-projections perpendicular to the
			beam direction can be reconstructed with a precision of about 3~$\mu$rad.
			
			Each RP is equipped with a stack of 10 silicon strip detectors designed with the specific objective of reducing the insensitive area at the edge facing the beam to only a few tens of micrometers. The 512 strips with 66
			$\mu$m pitch of each detector are oriented at an angle of +45$^{\circ}$ (five planes) and -45$^{\circ}$ (five planes) with respect to the
			detector edge~\cite{Ruggiero:2009zz}. The complete and detailed description of the TOTEM experiment is given in~\cite{Anelli:2008zza,TOTEM:2013iga}.
	\vspace{1mm}
	\vspace{2mm}

\section{Data taking and analysis}
The analysis has been performed on a large data sample, including seven data sets (DS1 - DS7) recorded in 2015, corresponding to the LHC fills 4495, 4496, 4499, 4505, 4509, 4510 and 4511, respectively.
The LHC beam was configured with the $\beta^{*}=90$~m optics described in detail in~\cite{Antchev:2011vs,Antchev:2013gaa,Nemes:2131667,Antchev:2014voa}.

The RP detectors were placed at a distance of 10 times the transverse beam size ($\sigma_{\rm beam}$) from the outgoing beams. The special trigger settings allowed 
to collect about $10^{9}$ elastic events.

The angular resolution is different for each of the data sets DS1-DS7, and it deteriorates with time within the fill, expected mainly due to the beam emittance growth according to $\sigma(x)=\sqrt{\varepsilon\beta}$~\cite{Niewiadomski:2008zz,Nemes:2131667}. The data sets have been reorganized according to their resolution into two larger
data sets. The ones with better(about 20~\%) resolution were collected into DS$_{\rm g}$, which includes DS1, DS2 and DS4. The remaining ones are collected in data set DS$_{\rm o}$. The statistical uncertainties of the scattering angles,
obtained from the data, are summarized in Table~\ref{resolutions} for the two data sets. 

The normalization of this analysis is based on the $\sqrt{s}=$13~TeV total cross-section measurement with $\beta^{*}=90$~m optics, where the RP detectors were placed two times closer (5$\sigma_{\rm beam}$ distance)
to the beam~\cite{Antchev:2017dia}. This data set ($DS_{\rm n}$) corresponds to the LHC fill 4489, recorded before DS1.

	\subsection{Elastic analysis}

	\subsubsection{Reconstruction of kinematics}
		The horizontal and vertical scattering angles of the proton at IP5 $(\theta_{x}^{*},\theta_{y}^{*})$  are reconstructed in a given arm by inverting the proton transport
		equations~\cite{Antchev:2014voa}
			\begin{align}
			    \theta_{x}^{*} = \frac{1}{\frac{{\rm d}L_{x}}{{\rm d}s}}\left(\theta_{x}-\frac{{\rm d} v_{x}}{{\rm d} s}x^{*}\right)\,,\,
    			    \theta^{*}_{y} = \frac{y}{L_{y}}\,,
			    \label{reconstruction_formula_theta_x_rearranged}
			\end{align}
			where $s$ denotes the distance from the interaction point, $y$ is the vertical coordinate of the proton's trajectory, $\theta_{x}$ is its horizontal
			angle at the detector, and $x^{*}$ is the horizontal vertex coordinate reconstructed as
			\begin{align}
			    x^{*}&=\frac{L_{x,{\rm far}}\cdot x_{\rm near} - L_{x,{\rm near}}\cdot x_{\rm far}}{d}\,,
			    \label{reconstruction_formula_x}
			\end{align}
			where $d=( v_{x,{\rm near}}\cdot L_{x,{\rm far}} -  v_{x,{\rm far}}\cdot L_{x,{\rm near}})$. The scattering angles obtained for the two arms are averaged 
			and the four-momentum transfer squared is calculated as
			\begin{align}
			    t=-p^{2}\theta^{*2}\,,
			    \label{reconstructed_t}
			\end{align}
		where $p$ is the LHC beam momentum and the scattering angle $\theta^{*}=\sqrt{{\theta_{x}^{*}}^{2} + {\theta_{y}^{*}}^{2}}$. Finally, the azimuthal angle is 
			\begin{align}
				\phi^{*}=\arctan\left(\frac{\theta_{y}^{*}}{\theta_{x}^{*}}\right)\,.
			    \label{phistar}
			\end{align}

		The coefficients $L_{x}$, $L_{y}$ and $v_{x}$ of Eq.~(\ref{reconstruction_formula_theta_x_rearranged}) and Eq.~(\ref{reconstruction_formula_x}) are optical functions of the LHC beam determined by the
		accelerator magnets between IP5 and the RP detectors, see Fig.~\ref{layout_LHC}. The $\beta^{*}=90$~m optics is designed with a large vertical effective length $L_{y}\approx~263$~m at the RPs placed at $220$~m from IP5. Since the horizontal
		effective length $L_{x}$ is close to zero at the RPs, its derivative ${{\rm d}L_{x}}/{{\rm d}s}\approx-0.6$ and the local angle $\theta_{x}$ is used instead. The different reconstruction formula in the vertical
		and horizontal plane in Eq.~(\ref{reconstruction_formula_theta_x_rearranged}) is also motivated by their different sensitivity to LHC magnet and beam perturbations.

	\subsubsection{RP alignment and beam optics}
	\label{RP_alignment}

	After applying the usual TOTEM alignment methods the residual misalignment is about 3.3~$\mu$m in the horizontal coordinate and about 110~$\mu$m in the vertical. When propagated to the
	reconstructed scattering angles, this leads to uncertainties about 1.11~$\mu$rad (horizontal angle) and 0.42~$\mu$rad (vertical angle)~\cite{Antchev:2016vpy,Antchev:2015zza}.

	The nominal optics has been updated from LHC magnet and current databases and
	calibrated using the observed elastic candidates of $DS_{\rm n}$. The calibrated optics has been used in the analysis of $DS_{\rm g}$ and $DS_{\rm o}$ exploiting the stability of the LHC optics. The uncertainties of the optical functions have been estimated with a Monte Carlo program applying the optics calibration procedure
	on a sophisticated simulation of the LHC beam and its perturbations. The obtained uncertainty is about 1.2~$\permil$ for ${{\rm d}L_{x}}/{{\rm d}s}$ and $2.1~\permil$ for $L_{y}$~\cite{Antchev:2014voa,Nemes:2131667}.

	\subsubsection{Event selection}

	The analysis is similar to the procedure performed for the measurement of the elastic cross section at several other LHC energies~\cite{Antchev:2013paa,Antchev:2016vpy,Antchev:2011vs,Antchev:2013iaa,DIS2017_proceedings,Paper_2p76}. 
	The measurement of the elastic events is based on the selection of events with the following topology in the RP detector system: a reconstructed track in the near and far
	vertical detectors on each side of the IP such that the elastic signature is satisfied in one of the two diagonals: left bottom and right top (Diagonal 1) or left top and right bottom (Diagonal 2).

	\begin{figure}[H]
		\centering
		\includegraphics[width=0.485\columnwidth,height=52mm]{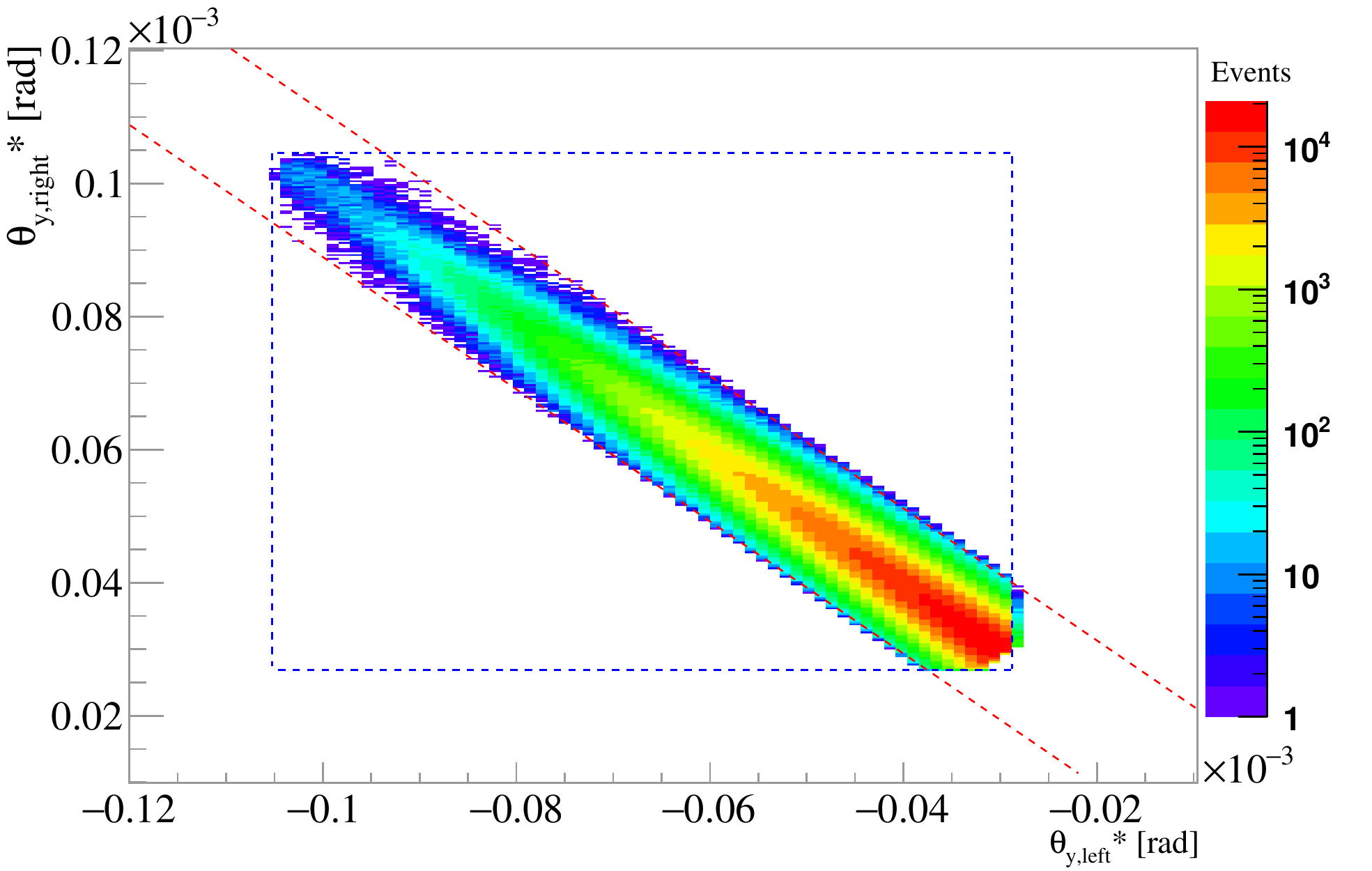}
		\includegraphics[width=0.485\columnwidth,height=52mm]{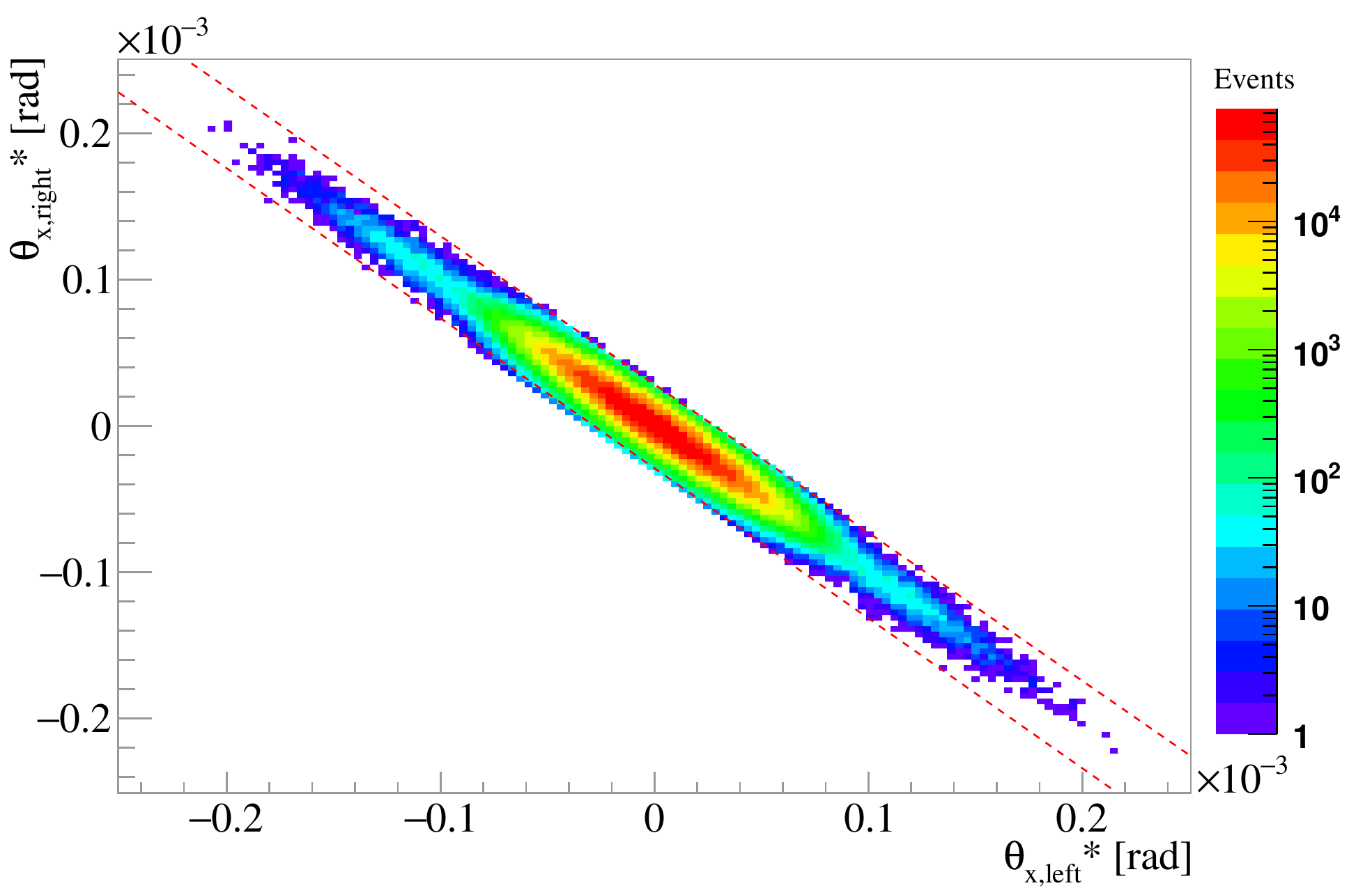}
		\caption{(color) The collinearity of the vertical and horizontal scattering angles. The blue lines represent the angular acceptance cuts in the vertical plane around the acceptance edges. The red ones show the 4$\sigma$ physics
		cuts to require the collinarity of the angles in both projections.}
		\label{cuts}
 	\end{figure}

	In addition, the elastic event selection requires the collinearity of the outgoing protons in the two arms, see Fig.~\ref{cuts}. The suppression of the
	diffractive events is also required using the correlation between the position $y$ and the inclination $\Delta y=y_{\rm far}-y_{\rm near}$ with so-called spectrometer cuts, see Table~\ref{cuts_table}. The
	equality of the horizontal vertex position $x^{*}$ reconstructed from the left and right arms is also required.

	\begin{table*}
        \begin{center}
            \caption{The physics analysis cuts and their characteristic width $\sigma$ for $DS_{\rm g}$ in Diagonal 1 (the other diagonal agrees within the quoted uncertainty). The width $\sigma$ of the horizontal and vertical collinearity cuts define the resolution in the scattering angle, see Fig.~\ref{cuts}.}
            \begin{tabular}{ | c | l l c c c |}
		\hline
                		& Name				& 		& $\sigma$	&	& \\\hline 
			1	& Vertical collinearity cut	& [$\mu rad$] 	& $1.87\pm0.01$ & 	&	\\ 
			2	& Spectrometer cut, left arm	& [$\mu$m] 	& $15.9\pm0.3$	& 	&	\\ 
			3	& Spectrometer cut, right arm	& [$\mu$m] 	& $14.6\pm0.3$	& 	&	\\ 
			4	& Horizontal vertex cut		& [$\mu$m] 	& $7.3\pm1.0$	& 	&	\\ 
			5	& Horizontal collinearity cut	& [$\mu rad$] 	& $4.96\pm0.02$	& 	&	\\\hline 
            \end{tabular}
        \label{cuts_table}
        \end{center}
    \end{table*}

\begin{figure}[H]
		\centering
		\includegraphics[width=0.85\columnwidth]{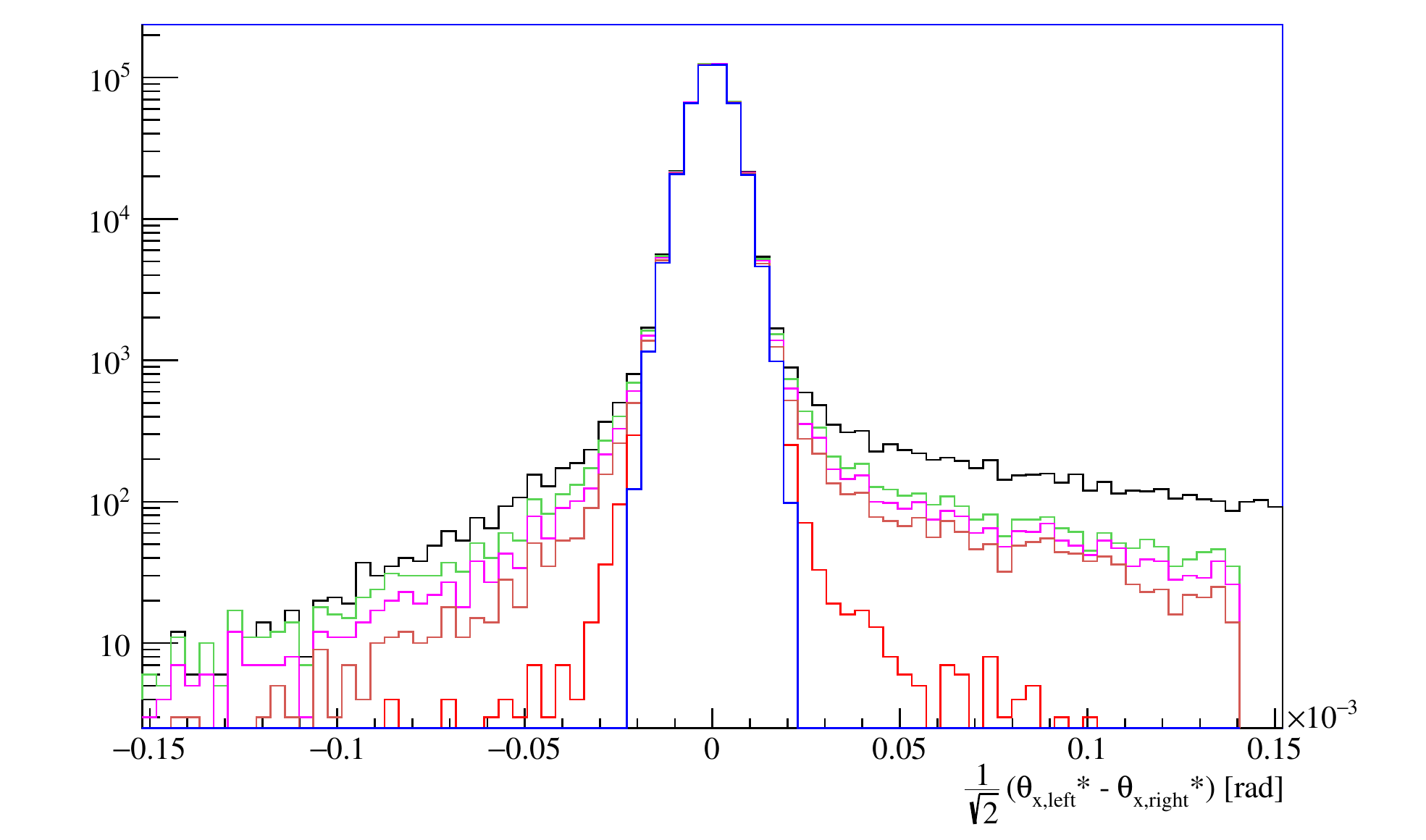}
		\caption{(color) The horizontal beam divergence estimated from the data of Diagonal 1 by comparing the reconstructed horizontal scattering angle $\theta_{x}^{*}$ of the left and right arm. The distribution is shown before
		any analysis cut (black solid line) and after each analysis cut following the order in Table~\ref{cuts_table}.}
		\label{signal_to_noise}
	\end{figure}

	Fig.~\ref{cuts} shows the horizontal collinearity cut imposing momentum conservation in the horizontal plane with 1~$\permil$ uncertainty. The cuts are applied at the 4$\sigma$ level, and they are optimized for
	purity (background contamination in the selected sample	less than 0.1~\%) and for efficiency (uncertainty of true elastic event selection 0.5~\%). Fig.~\ref{signal_to_noise} shows the progressive selection
	of elastic events after each analysis cut following the order in Table~\ref{cuts_table}.

	\subsubsection{Geometrical and beam divergence correction, unfolding}

	\begin{figure}[H]
		\centering
		\includegraphics[width=0.75\columnwidth]{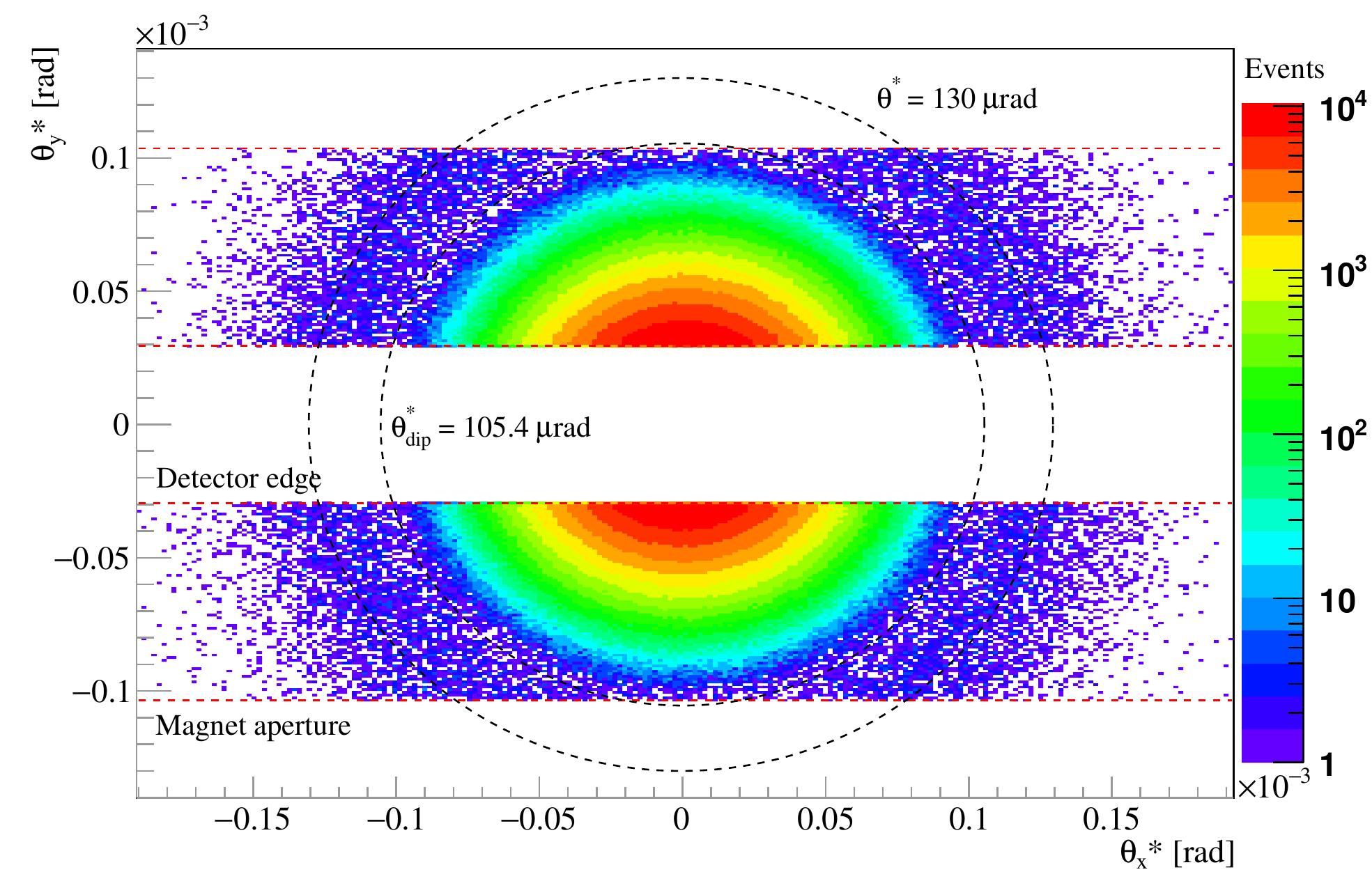}
		\caption{(color) The azimuthal distribution of the scattering angle $\theta^{*}$ demonstrates the azimuthal symmetry of elastic scattering on a data sample from Diagonal 1 and 2. The red dashed lines 
		show the analysis acceptance cuts, which define the acceptance boundaries near the detector edge and magnet aperture. The inner black dashed circle illustrates the approximate scattering angle	
		position $\theta_{\rm dip}^{*}$ of the diffractive minimum in the data.}
		\label{scattering_angle}
 	\end{figure}

	\begin{figure}
		\centering
		\includegraphics[width=0.85\columnwidth]{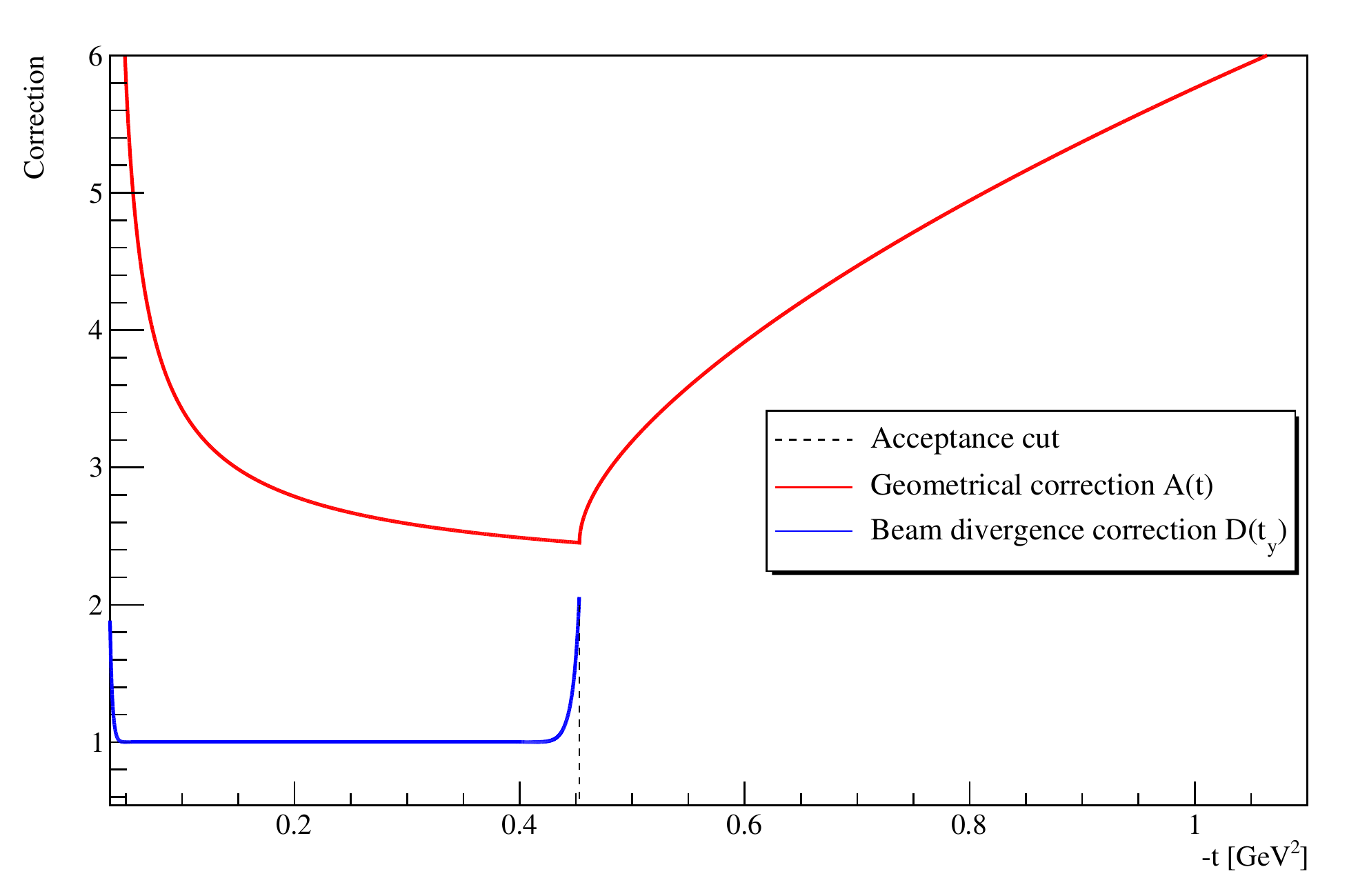}
		\caption{(color) The geometrical acceptance correction $\mathcal{A}(t)$ is defined by the cuts of Fig.~\ref{scattering_angle}. The beam divergence correction $D(t_{y})$ depends on the vertical collinearity 
		cut $\theta_{y}^{*}$ shown in Fig~\ref{cuts}: the angular window and the $\sigma$ of the cut determines the missing acceptance in the corners. Note that the data extends up to 4.0~GeV$^{2}$ due to the
		horizontal acceptance, see also Fig.~\ref{cuts}.}
		\label{acceptance_correction}
 	\end{figure}

		The acceptance of elastically scattered protons is limited by the RP silicon detector edge and by the LHC magnet apertures. The acceptance boundaries are defined by the
		acceptance cuts shown in Fig.~\ref{scattering_angle}. Fig.~\ref{scattering_angle} also shows the $|t|$-acceptance circle for the range of the diffractive minimum at $\theta^{*}_{\rm dip}=105.4~\mu$rad.

		The geometrical acceptance correction is calculated in order to correct for the missing acceptance in $\phi^{*}$ 
		\begin{equation}
			\mathcal{A}(t)=\frac{2\pi }{\Delta \phi^{*}(t)}\,,
			\label{acceptance_correction}
		\end{equation}
 		where $\Delta \phi^{*}$ is the visible azimuthal angle range.

		The correction function $\mathcal{A}(t)$ is drawn as a solid red line in Fig.~\ref{acceptance_correction}. The function is monotonically decreasing between the detector edge cut
		at $|t|_{y, \rm min}={ 0.04}$~GeV$^{2}$ and the aperture cut at $|t|_{y, \rm max}={ 0.45}$~GeV$^{2}$, since in this range the visible $\phi$ part of the acceptance circles is increasing with $|t|$.
		At the largest $|t|$-values of the analysis, about 4~GeV~$^{2}$, the maximum of the correction function $\mathcal{A}(t)$ is about 13.

		The acceptance cut at the acceptance edges is not step function-like due to the angular spread, the beam divergence, of the LHC beam. The effect of the beam divergence can be directly measured by comparing
		the angles reconstructed from the left and right arm, see Fig.~\ref{cuts}. The figure shows the correlation between the angles and at the same time the spread due to the beam divergence. The figure also shows the
		missing corners of the acceptance at the acceptance edges. This additional acceptance loss is modeled with a Gaussian distribution, with experimentally determined parameters.
		The beam divergence correction $\mathcal{D}(t_{y})$ is drawn as a solid blue line in Fig.~\ref{acceptance_correction}, which is close to 1 except at the acceptance edges.

		Finally, the acceptance correction $\mathcal{A}(t)\,\mathcal{D}(t_{y})$ is factorized in terms of the geometrical and beam divergence corrections.

	\begin{table*}\small
        \begin{center}
            \caption{Horizontal and vertical angular resolutions of the analysis data sets DS$_{\rm g}$ and DS$_{\rm o}$, respectively.}
            \begin{tabular}{ | c | c  c |}
		\hline
                		 		& Horizontal			& Vertical					\\ 
						& \multicolumn{2}{c|}{[$\mu$rad]} \\ \hline
			DS$_{\rm g}$		& 4.96\,$\pm$\,0.02		& 1.87\,$\pm$ 0.01				\\ 
			DS$_{\rm o}$		& 5.10\,$\pm$\,0.02		& 2.24\,$\pm$ 0.01				\\\hline 
            \end{tabular}
        \label{resolutions}
        \end{center}
    \end{table*}

		The unfolding of resolution effects has been estimated with a Monte Carlo simulation whose parameters are obtained from the data, see Table~\ref{resolutions}. The probability distribution $p(t)$ of the
		event generator is based on the fit of the differential rate ${\rm d}N/{\rm d}t$. Each generated MC event is propagated to the RP detectors with the proper model of the LHC optics, which takes
		into account the beam divergence and other resolution effects. The kinematics of the event is reconstructed and a histogram is built from
		the four momentum transfer squared $t$ values. The ratio of the histograms without and with resolution effect describes the first approximation of the bin-by-bin corrections due to bin migration. The probability distribution
		$p(t)$ of the simulation is multiplied with the correction histogram, to modulate the source, and the procedure is repeated until the histogram with migration effects
		coincides with the measured distribution, thus the correct source distribution has been found. The uncertainty of the unfolding procedure is estimated from the residual difference between the measured histogram
		${\rm d}N_{\rm el}/{\rm d}t$ and the simulated histogram with resolution effects.

\begin{figure}
		\centering
		\includegraphics[width=0.48\columnwidth,height=52mm]{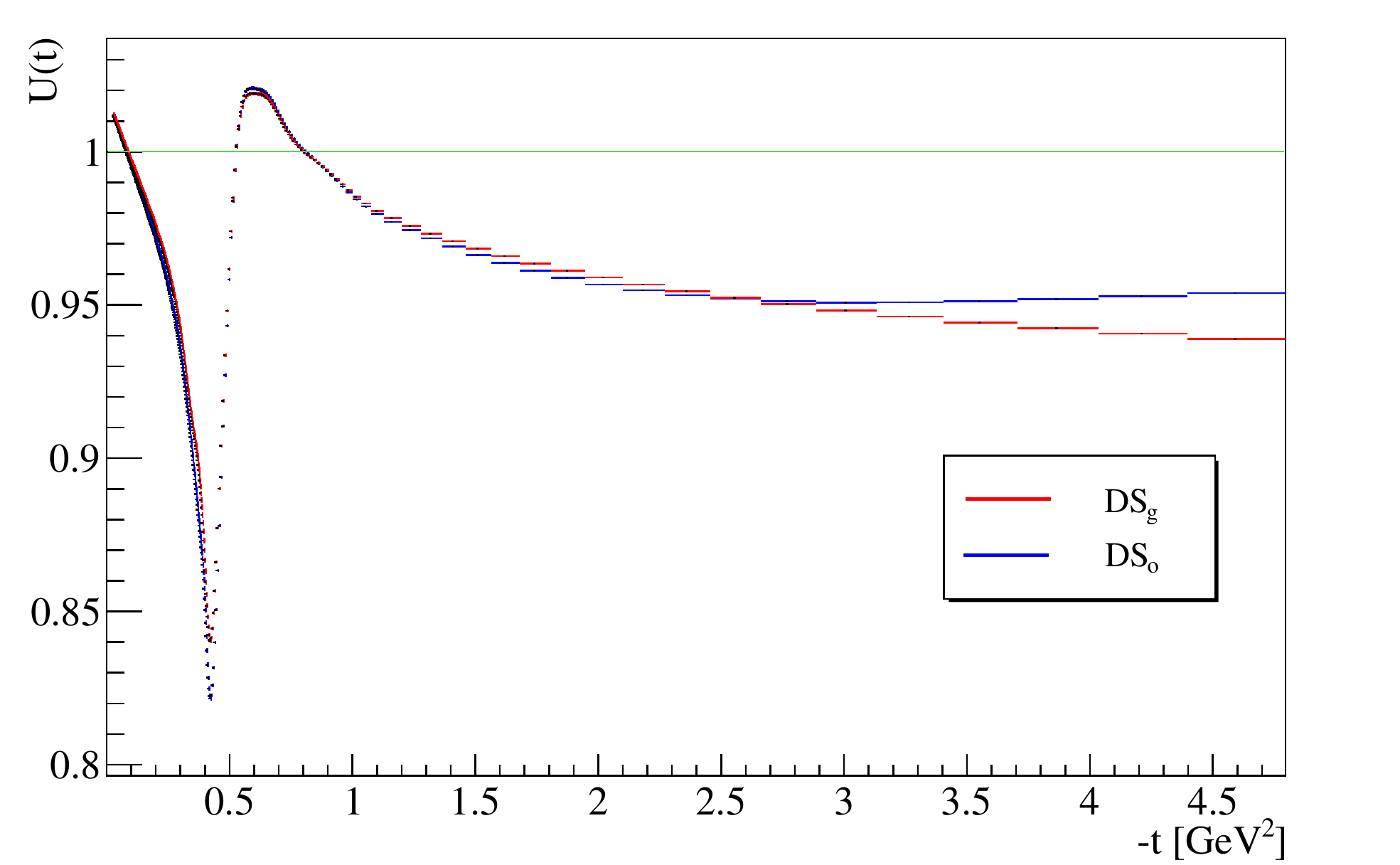}
		\includegraphics[width=0.48\columnwidth,height=52mm]{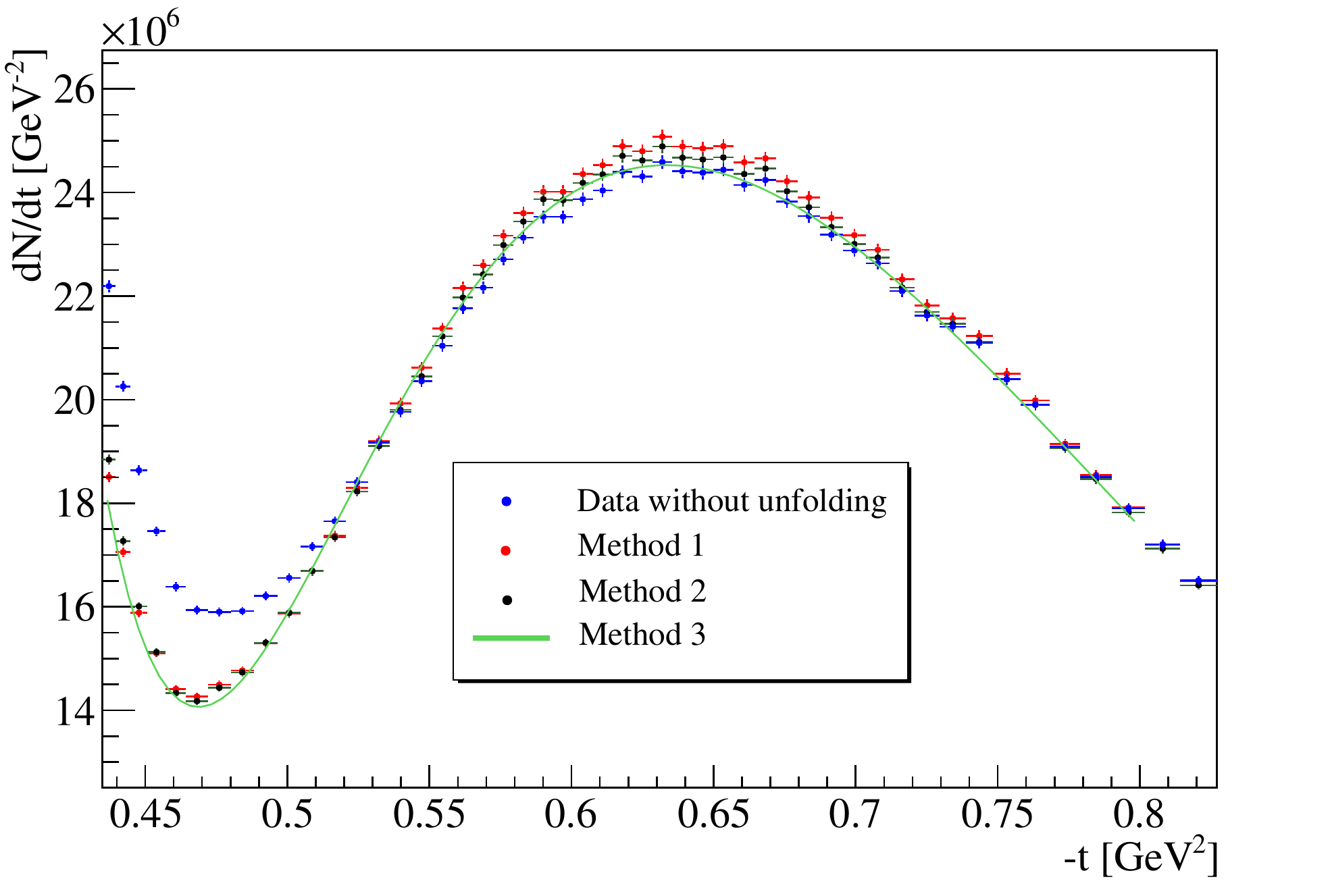}
		\caption{(color) 
		The unfolding correction histograms $U(t)$ for data set DS$_{g}$ and DS$_{o}$, which correspond to the resolution values summarized in Table~\ref{resolutions} (left panel).
		Comparison of the unfolding of ${\rm d}N_{\rm el}/{\rm d}t$ using the three unfolding methods around the diffractive minimum (right panel).}
		\label{comparison_of_unfolding}
 	\end{figure}

		According to Table~\ref{resolutions} the angular resolution
		is different in the horizontal and vertical plane, so the simulation takes into account the angular acceptance cuts of the analysis to give the proper weight to the resolution effects. The angular spread of
		the beam is determined with an uncertainty of 0.1 $\mu$rad by comparing the scattering angles reconstructed from the left and right arm, therefore the unfolding correction factor $\mathcal{U}(t)$ can
		be calculated with a precision better than 0.1~\%. The unfolding correction histograms $U(t)$ are shown in Fig.~\ref{comparison_of_unfolding}. Three different unfolding methods have been compared in order to
		estimate the contribution of the unfolding to the systematic uncertainty: the described MC based algorithm (Method 1), regularized unfolding (Method 2)
		and deconvolution of a proper fit function with resolution $\sigma$ (Method 3)~\cite{Cowan_unfolding}. The results of the three methods are perfectly consistent within their uncertainties, see Fig.~\ref{comparison_of_unfolding}.

		In total, the event-by-event correction factor due to acceptance corrections and resolution unfolding is
		\begin{equation}
			\mathcal{C}(t,t_{y})=\mathcal{A}(t)\,\mathcal{D}(t_{y})\,\mathcal{U}(t)\,.
		\end{equation}

	\subsubsection{Inefficiency corrections}
		{\color{black} The proton reconstruction efficiency of the RP detectors is evaluated directly from the data. The RP detectors are unable to resolve multiple tracks, which is the main source
		of detector inefficiency~\cite{Anelli:2008zza}. The additional tracks can be due to interactions of the protons with the sensors or the surrounding material, or due to pileup with non-signal protons or beam halo.

		\begin{figure}[H]
			\centering
			\includegraphics[width=0.75\columnwidth]{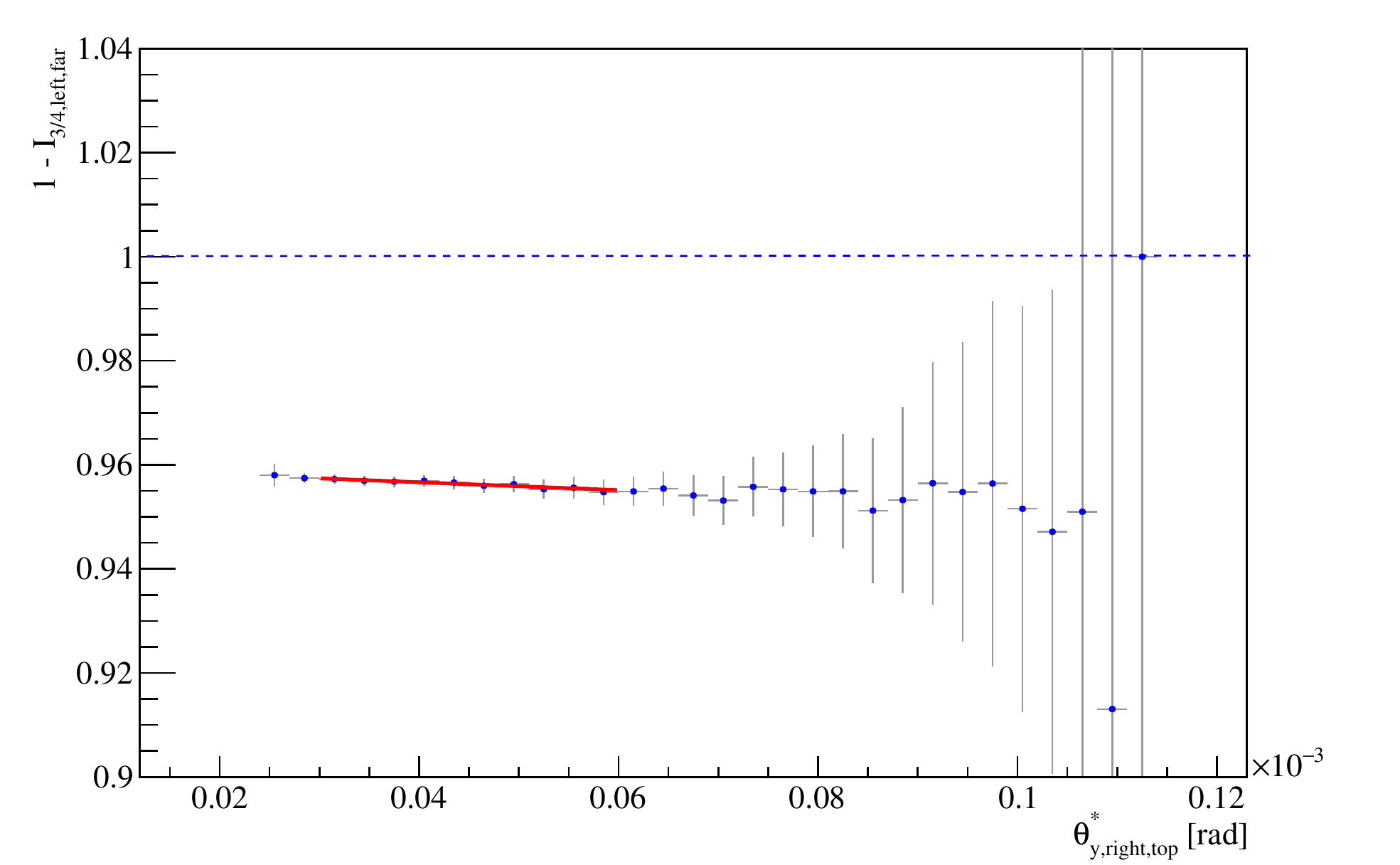}
			\caption{(color) The contribution of the left far pot to the $\mathcal{I}_{\rm 3/4}$ correction in Diagonal 1. The inefficiency as a function of $\theta_{y}^{*}$ is estimated with a 
			linear fit, shown with a solid red line.}
			\label{the_3_4_correction}
 		\end{figure}

    \begin{table*}
        \begin{center}
            \caption{Corrections to the differential rate for the two diagonals. The ``uncorrelated'' inefficiency correction ($\mathcal{I}_{\rm 3/4}$)
		is $t$-dependent, in the table the average correction on the elastic rate is provided.}
            \begin{tabular}{ | c  c  c  |}
		\hline
				Correction			 	 & Diagonal 1				& Diagonal 2					\\\hline\hline 
			$\mathcal{I}_{\rm 3/4}$~[\%]	 		 & $11.2~\pm~1.1$			& $12.25~\pm~1.4$				\\ \hline 
            \end{tabular}
        \label{analysis_corrections_summary}
        \end{center}
    \end{table*}

		The inefficiency corrections are calculated for different categories: ``uncorrelated'' ($\mathcal{I}_{\rm 3/4}$), when one RP out of four along a diagonal
		has no reconstructed track; this inefficiency includes the loss of the elastic proton due to an additional track coming from nuclear interaction, shower or pile-up with a beam halo proton. The $\mathcal{I}_{\rm 3/4}$
		inefficiency has been determined using a reduced set of elastic cuts in a so-called ``3/4``  elastic analysis per detector and diagonal~\cite{Antchev:2015zza}. The inefficiency
		is determined as a function of $\theta_{y}^{*}$ per RP, see Fig.~\ref{the_3_4_correction}, which shows that the dependence on the angle is close to negligible. The overall correction
		on the elastic rate is described in Table~\ref{analysis_corrections_summary}.

 		The inefficiency is called ``correlated'' ($\mathcal{I}_{\rm 2/4}$) when both RP of one arm have no reconstructed tracks. The case when the inefficient RPs are in different arms is denoted with $\mathcal{I}_{\rm 2/4\,diff.}$. The
		present analysis focuses on the differential cross-section measurement, and its overall normalization is determined from the corresponding cross section analysis at 13~TeV~\cite{Antchev:2017dia}. The $t$-dependence of the inefficiencies $\mathcal{I}_{\rm 2/4}$ and  $\mathcal{I}_{\rm 2/4\,diff.}$ 
		is even weaker than for $\mathcal{I}_{\rm 3/4}$, thus these inefficiencies are estimated but set to zero in the total correction factor per event (shown in its most general form)
		\begin{equation}
			f(t,t_{y})=\frac{1}{\eta_{\rm d}\eta_{\rm tr}}\cdot\frac{\mathcal{C}(t,t_{y})}{1-\mathcal{I}}\cdot\frac{1}{\Delta t}\,,
			\label{correction}
		\end{equation}
		where the track reconstruction inefficiency $\mathcal{I}=\mathcal{I}_{3/4}(\theta_{y}^{*}) + \mathcal{I}_{\rm 2/4}+\mathcal{I}_{\rm 2/4\,diff}=\mathcal{I}_{3/4}(\theta_{y}^{*})$ and
		$\Delta t$ is the bin width. The $\eta_{\rm d}$, $\eta_{\rm tr}$ are the DAQ and trigger efficiencies that influence the normalization only. However, during the final normalization
		to the total cross-section these parameters cancel~\cite{Antchev:2017dia}.

\section{The differential cross section}
	After inefficiency correction the differential rates ${\rm d}N/{\rm d}t$ of the two diagonals (Diagonal 1 and 2) agree
	within their statistical uncertainty. The two diagonals are almost independent measurements, thus the
	final measured differential rate is calculated as the bin-by-bin weighted average of the two differential
	elastic rates ${\rm d}N_{\rm el}/{\rm d}t$, according to their systematic uncertainty.

	The normalization is based on the 13~TeV total cross-section measurement with $\beta^{*}=90$~m optics, where the RP detectors were placed at half the distance to the beam (5$\sigma_{\rm beam}$ instead of 10$\sigma$ distance)~\cite{Antchev:2017dia}. The $\rho$ measurement at 13~TeV with $\beta^{*}=2500$~m optics was also essential to obtain the final normalization~\cite{Antchev:2298154}. The 
	differential cross-section is shown in Fig.~\ref{differential_cross_section}. The normalization uncertainty $5.5$~\% is determined by the total cross-section measurement,
	inheriting the normalization uncertainty from~\cite{Antchev:2017dia}.

	The numerical values of the differential cross-section, the representative $|t|$-values, as well as the statistical and systematic uncertainties are given in Tables 5-10.
	
	\begin{figure*}[h]
		\centering
		\includegraphics[width=1.0\linewidth]{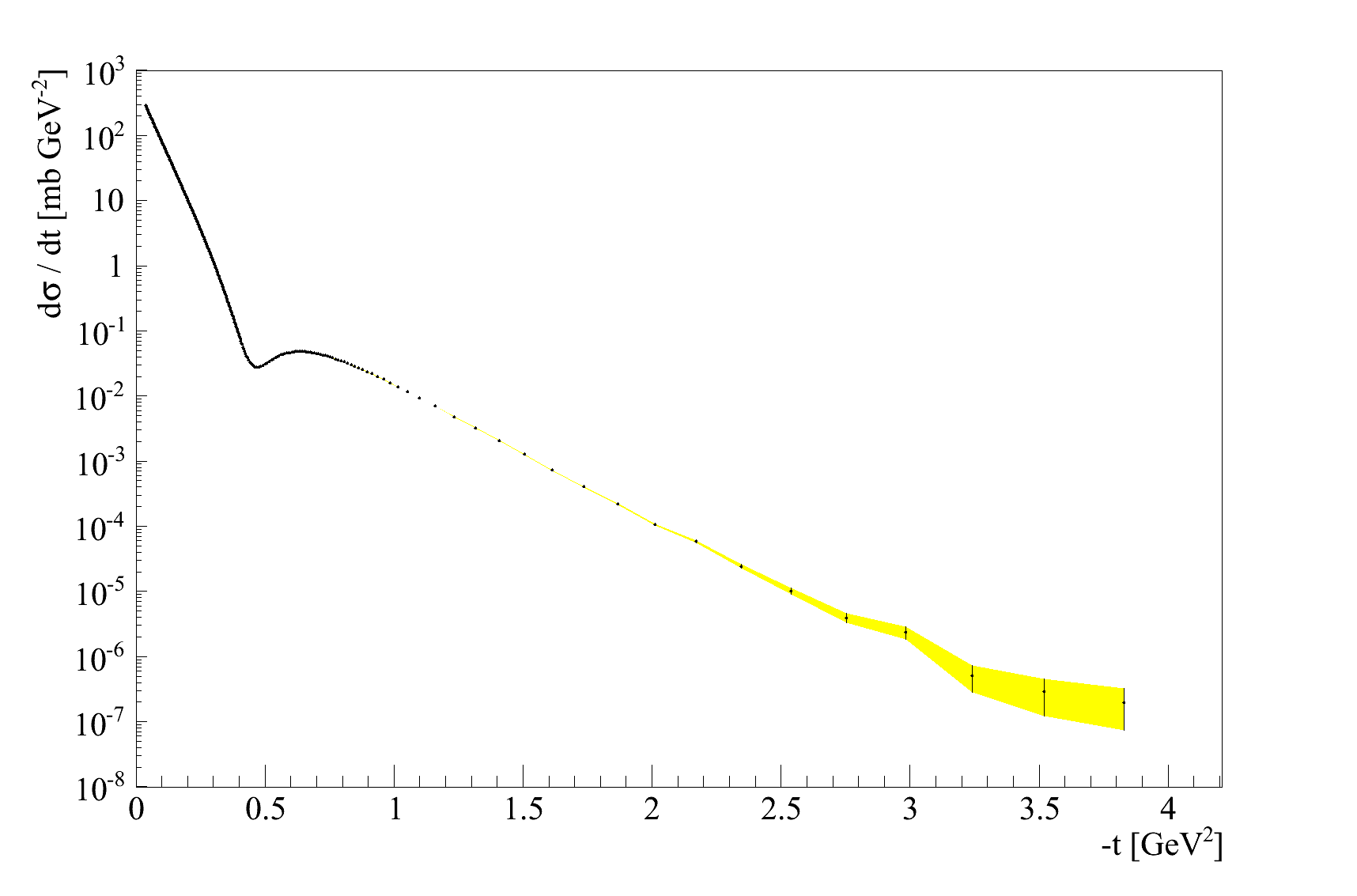}
		\caption{(color) Differential elastic cross-section ${\rm d}\sigma/{\rm d}t$ at $\sqrt{s} = 13$~TeV. The statistical and  $|t|$-dependent correlated systematic uncertainty envelope is shown
		as a yellow band.}
		\label{differential_cross_section}
	\end{figure*}

	The propagation of systematic uncertainties to the $|t|$-distribution has been estimated with a Monte Carlo program, see Fig.~\ref{error_band_of_syst}. A fit of the final differential cross-section data is used to 
	generate the true reference $|t|$-distribution. Simultaneously, another $|t|$-distribution is created, which is perturbed with one of the systematic effects at $1\sigma$ level.
	The difference between the $|t|$-distributions gives the systematic effect on the differential cross-section
	\begin{align}
	 \delta s_{q}(t)\equiv \frac{\partial({\rm d}\sigma/{\rm d}t)}{\partial q}\delta q\,,
	 \label{estimation_of_systematics}
	\end{align}
	where $\delta q$ corresponds to $1\sigma$ bias in the quantity $q$ responsible for a given systematic effect. The Monte-Carlo simulations show
	that the combined effect of several systematic errors is well approximated by linear combination of the individual contributions from Eq.~(\ref{estimation_of_systematics}).
	\begin{figure}[H]
		\centering
		\includegraphics[width=0.9\columnwidth]{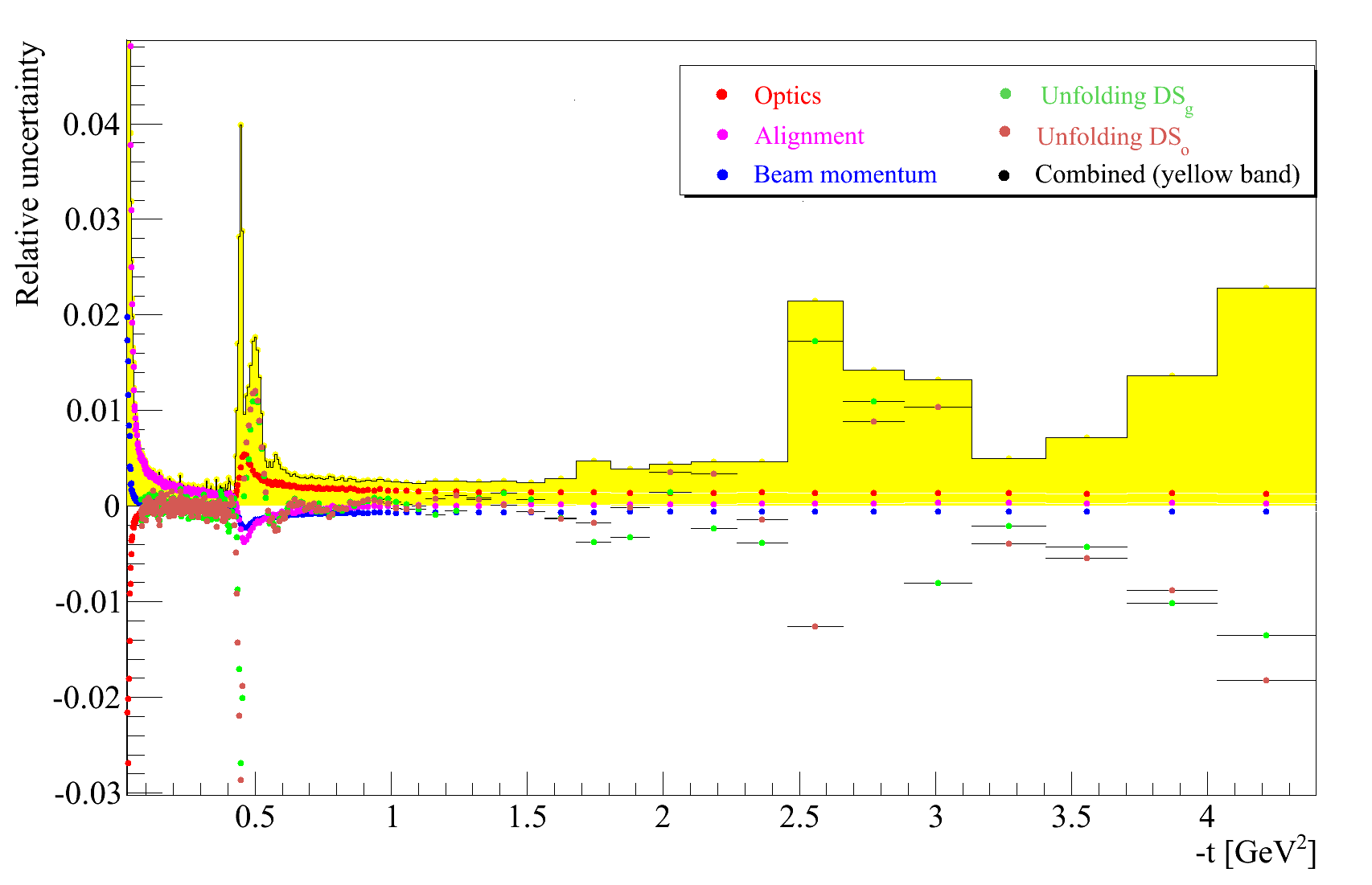}
		\caption{(color) Summary of the $|t|$-dependent systematic uncertainties. The figure shows the systematic uncertainties due to $1\sigma$ optics, alignment and beam momentum perturbations. The
		contribution of the unfolding into the systematic uncertainty is also presented. The yellow band is the combined systematics uncertainty.}
		\label{error_band_of_syst}
 	\end{figure}

	The $|t|$-dependent systematic uncertainties are summarized in Fig.~(\ref{error_band_of_syst}). The result can be used to approximate the covariance matrix of systematic uncertainties:
	\begin{align}
		V_{ij}=\sum_{q}\delta s_{q}(i)\delta s_{q}(j)\,,
	 \label{covariance_matrix}
	\end{align}
	where $i$ and $j$ are bin indices, and the sum over $q$ goes over the optics, alignment and beam momentum error contributions. The model fits of the data have been evaluated using
	the covariance matrix in the generalized least-squares method
	\begin{align}
		\chi^{2}=\Delta^{T}V^{-1}\Delta,\quad\Delta_{i}=\left(\frac{{\rm d}\sigma}{{\rm d}t}-\left.f(t)\right|_{t=t_{{\rm repr}}}\right)_{{\rm bin}\, i}\,
	 \label{chi2_with_correlations}
	\end{align}
	and $V=V_{\rm stat} + V_{\rm syst}$.

	\begin{figure*}
		\includegraphics[width=1.0\linewidth]{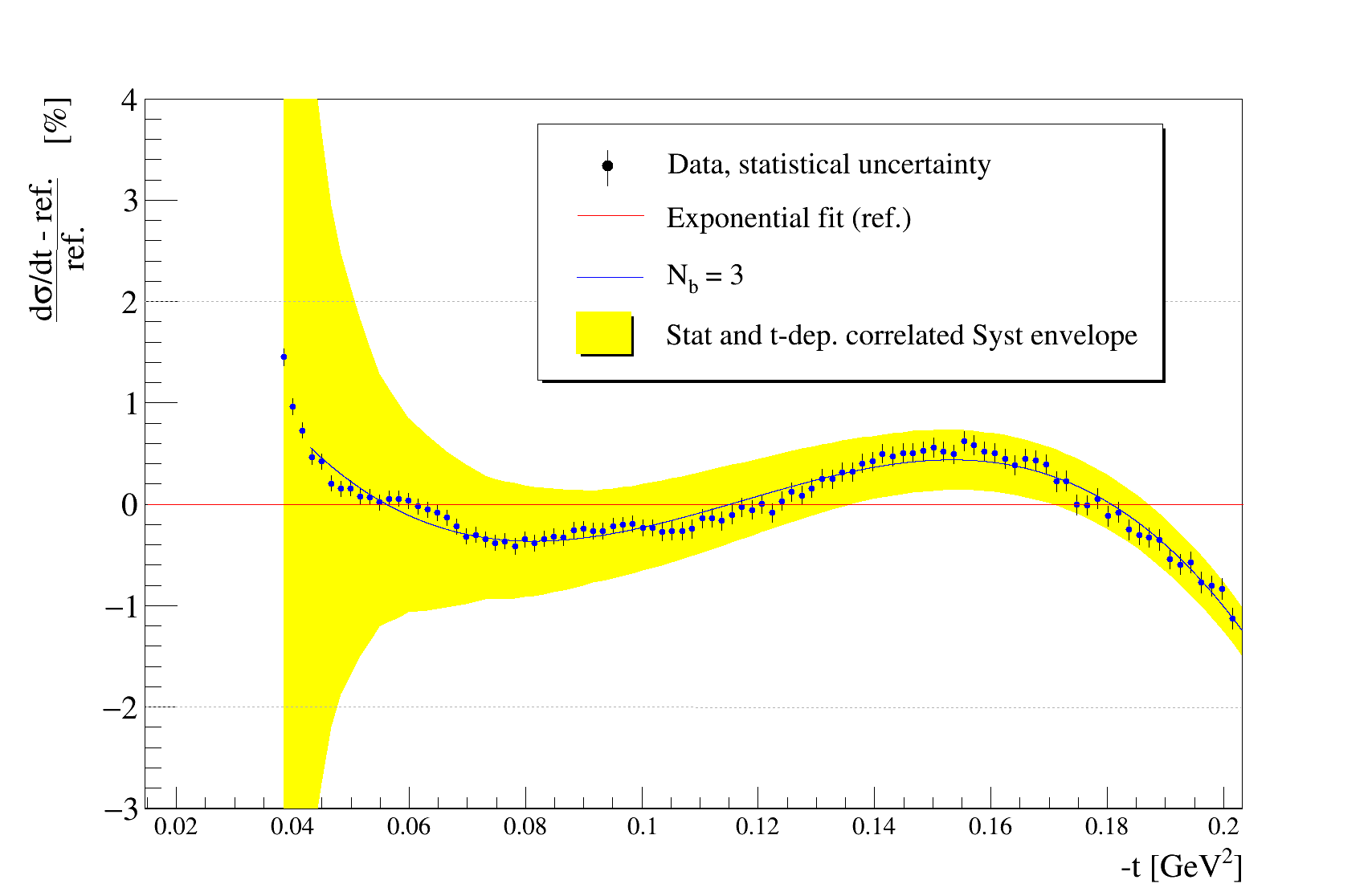}
		\caption{The non-exponential part of the data. The statistical and  $|t|$-dependent correlated systematic uncertainty envelope is shown as a yellow band, while the data points show the statistical uncertainty.}
		\label{differential_cross_section3}
	\end{figure*}

	The nuclear slope has been found to be $B=(20.40 \pm 0.002^{\rm stat} \pm 0.01^{\rm syst})~$GeV$^{-2}$ using an exponential fit in the $|t|$-range from
	$0.04~$GeV$^{2}$ to $0.2~$GeV$^{2}$. The relative difference between data and this fit is plotted in Fig.~\ref{differential_cross_section3}, showing a non-exponential shape, similar to the 8~TeV result~\cite{Antchev:2015zza}. Consequently, the value
	found for the nuclear slope $B$ can be considered as an average $B$ and the fit quality $\chi^{2}/{\rm ndf}=1175.3/92$ shows that the exponential model is an
	oversimplified description of the data. To obtain a better fit one can generalize the pure exponential
	to a cumulant expansion: 
	\begin{align}
		\frac{{\rm d}\sigma}{{\rm d}t}(t)=\left.\frac{{\rm d}\sigma}{{\rm d}t}\right|_{t=0}\exp\left(\sum_{i=1}^{N_{b}}b_{i}t^{i}\right)\,,
	 	\label{exponential_like}
	\end{align}

	where the $N_{b}=1$ case corresponds to the exponential. The $N_{b}=3$ case is the first which provides a satisfactory description of the data, with $\chi^{2}/{\rm ndf}=109.5/90$ and $p\scalebox{0.85}{-}{\rm value}$=8.0~\%, see Fig.~\ref{differential_cross_section3}.

	The diffractive minimum and the subsequent maximum has been observed with great accuracy, see Fig.~\ref{differential_cross_section1}. The dip position has been found to be $|t_{\rm dip}|=(0.47 \pm 0.004^{\rm stat} \pm 0.01^{\rm syst})~$GeV$^{2}$. The
	statistical uncertainty is the half the bin width, while the systematic is determined by the combined $|t|$-resolution of the two diagonals. The ratio of the differential cross-section values at the diffractive minimum and at the subsequent
	maximum has been found to be $R=1.77\pm0.01$. The value of $R$ is calculated from the value of the maximum and minimum bin, since the data is very precise and the bin-by-bin fluctuations are on the $5~\permil$ level.
	The uncertainty is calculated from the statistical uncertainty of the two bins, since the systematic uncertainty of the dip and bump follows the same pattern, hence the systematic uncertainty of $R$ is negligible.

	The large-$|t|$ part of the measured differential cross section, starting from $t=$2.1~GeV$^{2}$ up to 4~GeV$^{2}$, is consistent with a power law behavior ($p\scalebox{0.85}{-}{\rm value}$=80~\%). The fitted
	exponent is of the order of $10$, compatible with lower energy measurements~\cite{Amaldi:1979kd}, and high energy predictions~\cite{Brodsky:1973kr}.

	\begin{figure*}[h]
		\centering
		\includegraphics[width=1.0\linewidth]{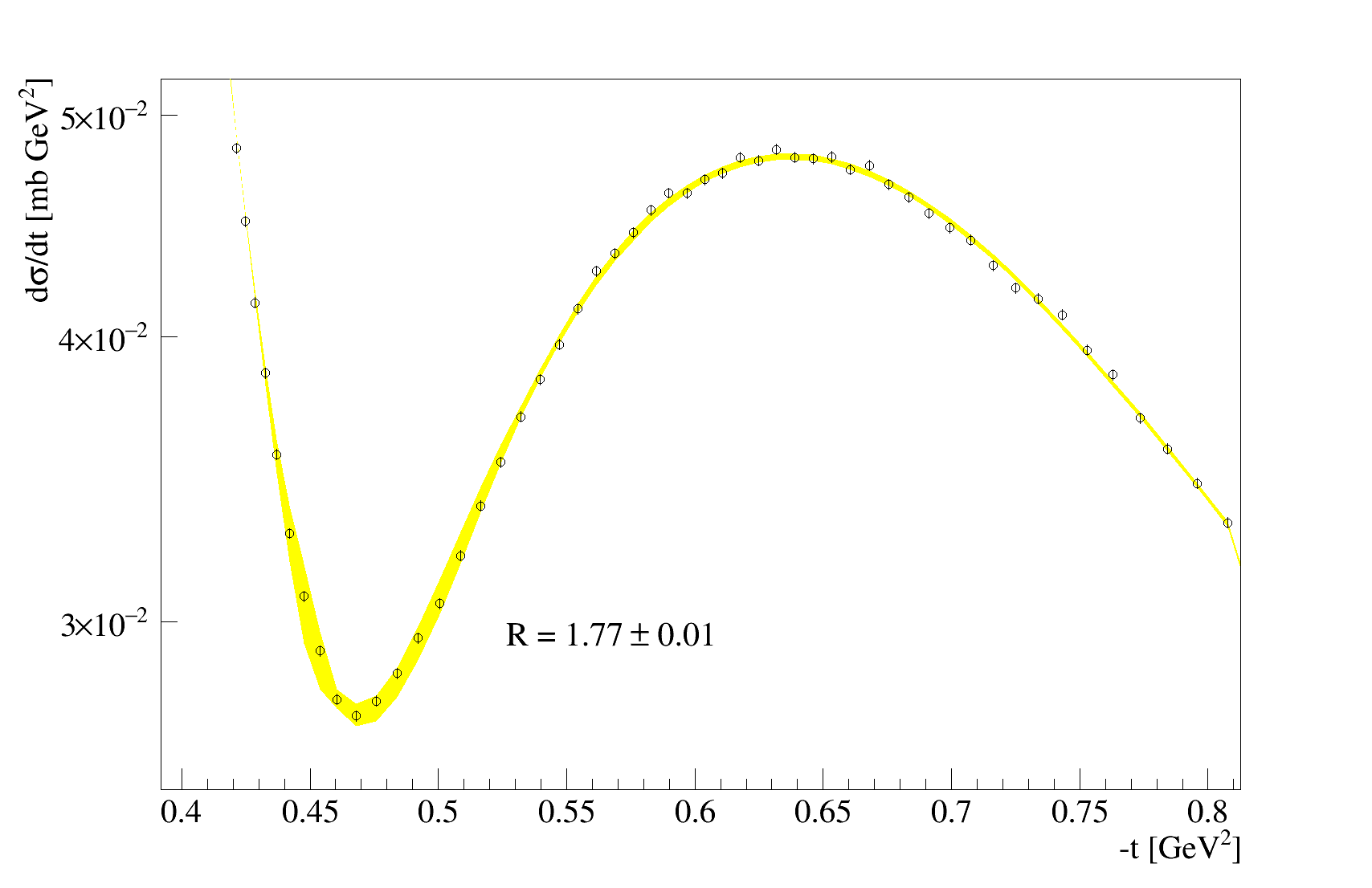}
		\caption{The diffractive minimum has been observed with high significance at 13~TeV. The uncertainty on the points is the statistical uncertainty, while the yellow band
		shows the full uncertainty, including the systematic part. The dip position has been found to be $|t_{\rm dip}|=(0.47 \pm 0.004^{\rm stat} \pm 0.01^{\rm syst})~$GeV$^{2}$ and the
		differential cross section ratio between the second maximum and the minimum is $R=1.77\pm0.01^{\rm stat}$.}
		\label{differential_cross_section1}
	\end{figure*}

	\begin{table*}[h]
		\centering
		\caption{The main physics observables and their statistical and systematic uncertainty.}
            \begin{tabular}{| c  c  c  c |}
		\hline
                	Physics quantity			& \multicolumn{2}{c}{Value} 		&  Total uncertainty 	\\ 
                	\multicolumn{3}{|c}{} 									&  Stat. $\oplus$ Syst. 	\\ \hline
			$B$~[GeV$^{-2}$]			& \multicolumn{2}{c}{20.40}			& $0.002\oplus0.01$= 0.01\\
			$|t_{\rm dip}|$~[GeV$^{2}$]		& \multicolumn{2}{c}{0.47}			& $0.004\oplus0.01$= 0.01\\ 
			$R$					& \multicolumn{2}{c}{1.77}			& $0.01$\\\hline 
            \end{tabular}
		\label{physics_quantities}
	\end{table*}

\section{Summary}
The TOTEM collaboration has measured the elastic proton-proton differential cross section ${\rm d}\sigma/{\rm d}t$ at $\sqrt{s}=13$~TeV LHC energy in the four-momentum transfer squared $|t|$ range
from $0.04$~GeV$^{2}$ to $4$~GeV$^{2}$. The special data acquisition allowed to collect about 10$^{9}$ elastic events and the precise measurement
of the differential cross-section including, the diffractive minimum and the large-$|t|$ tail. The average nuclear slope has been found to be $B=(20.40 \pm 0.002^{\rm stat} \pm 0.01^{\rm syst})~$GeV$^{-2}$ in the $|t|$-range
$0.04~$GeV$^{2}$ to $0.2~$GeV$^{2}$. The position of the diffractive minimum is $|t_{\rm dip}|=(0.47 \pm 0.004^{\rm stat} \pm 0.01^{\rm syst})~$GeV$^{2}$ and the differential cross section ratio at the maximum and
minimum is $R=1.77\pm0.01^{\rm stat}$ with negligible systematic uncertainty.

    \begin{table*}\small\color{black}
        \begin{center}
            \caption{The differential cross-section ${\rm d}\sigma/{\rm d}t$}
            \begin{tabular}{ | c  c  c  c  c  c|}
		\hline
			$|t|_{\rm low}$		& $|t|_{\rm high}$		&	$|t|_{\rm repr.}$		&	${\rm d}\sigma/{\rm d}t$	& Statistical uncertainty 	&	Systematic uncertainty \\ 
						& [GeV$^{2}$]			&					&					&	[mb GeV$^{-2}$]		& \\ \hline 

		0.03763	 & 0.03926	 & 0.03840	 & 291.005	 & 0.238	 & 23.23	 \\
		0.03926	 & 0.04090	 & 0.04004	 & 280.102	 & 0.219	 & 17.10	 \\
		0.04090	 & 0.04254	 & 0.04168	 & 270.253	 & 0.208	 & 13.45	 \\
		0.04254	 & 0.04419	 & 0.04332	 & 260.682	 & 0.198	 & 10.17	 \\
		0.04419	 & 0.04583	 & 0.04496	 & 251.980	 & 0.191	 & 8.04	 \\
		0.04583	 & 0.04748	 & 0.04661	 & 243.144	 & 0.183	 & 6.23	 \\
		0.04748	 & 0.04912	 & 0.04825	 & 234.997	 & 0.177	 & 5.10	 \\
		0.04912	 & 0.05077	 & 0.04990	 & 227.243	 & 0.171	 & 4.35	 \\
		0.05077	 & 0.05242	 & 0.05155	 & 219.549	 & 0.165	 & 3.65	 \\
		0.05242	 & 0.05408	 & 0.05320	 & 212.265	 & 0.159	 & 3.08	 \\
		0.05408	 & 0.05573	 & 0.05486	 & 205.123	 & 0.154	 & 2.55	 \\
		0.05573	 & 0.05739	 & 0.05651	 & 198.390	 & 0.149	 & 2.27	 \\
		0.05739	 & 0.05904	 & 0.05817	 & 191.804	 & 0.144	 & 2.01	 \\
		0.05904	 & 0.06070	 & 0.05983	 & 185.387	 & 0.139	 & 1.77	 \\
		0.06070	 & 0.06236	 & 0.06149	 & 179.122	 & 0.135	 & 1.63	 \\
		0.06236	 & 0.06402	 & 0.06315	 & 173.102	 & 0.130	 & 1.49	 \\
		0.06402	 & 0.06569	 & 0.06481	 & 167.281	 & 0.126	 & 1.36	 \\
		0.06569	 & 0.06735	 & 0.06648	 & 161.621	 & 0.122	 & 1.24	 \\
		0.06735	 & 0.06902	 & 0.06814	 & 156.083	 & 0.118	 & 1.13	 \\
		0.06902	 & 0.07069	 & 0.06981	 & 150.719	 & 0.114	 & 1.03	 \\
		0.07069	 & 0.07236	 & 0.07148	 & 145.698	 & 0.110	 & 0.94	 \\
		0.07236	 & 0.07403	 & 0.07315	 & 140.756	 & 0.107	 & 0.85	 \\
		0.07403	 & 0.07571	 & 0.07482	 & 135.985	 & 0.103	 & 0.81	 \\
		0.07571	 & 0.07738	 & 0.07650	 & 131.443	 & 0.100	 & 0.76	 \\
		0.07738	 & 0.07906	 & 0.07818	 & 126.963	 & 0.097	 & 0.72	 \\
		0.07906	 & 0.08074	 & 0.07985	 & 122.778	 & 0.094	 & 0.67	 \\
		0.08074	 & 0.08242	 & 0.08153	 & 118.598	 & 0.091	 & 0.63	 \\
		0.08242	 & 0.08410	 & 0.08322	 & 114.650	 & 0.088	 & 0.60	 \\
		0.08410	 & 0.08579	 & 0.08490	 & 110.811	 & 0.085	 & 0.56	 \\
		0.08579	 & 0.08747	 & 0.08658	 & 107.055	 & 0.082	 & 0.53	 \\
		0.08747	 & 0.08916	 & 0.08827	 & 103.514	 & 0.080	 & 0.49	 \\
		0.08916	 & 0.09085	 & 0.08996	 & 100.023	 & 0.077	 & 0.46	 \\
		0.09085	 & 0.09254	 & 0.09165	 & 96.617	 & 0.075	 & 0.43	 \\
		0.09254	 & 0.09423	 & 0.09334	 & 93.343	 & 0.072	 & 0.41	 \\
		0.09423	 & 0.09593	 & 0.09503	 & 90.220	 & 0.070	 & 0.39	 \\
		0.09593	 & 0.09762	 & 0.09673	 & 87.164	 & 0.068	 & 0.38	 \\
		0.09762	 & 0.09932	 & 0.09842	 & 84.210	 & 0.066	 & 0.36	 \\
		0.09932	 & 0.10102	 & 0.10012	 & 81.310	 & 0.064	 & 0.34	 \\
		0.10102	 & 0.10272	 & 0.10182	 & 78.544	 & 0.062	 & 0.33	 \\
		0.10272	 & 0.10442	 & 0.10353	 & 75.835	 & 0.069	 & 0.31	 \\
		0.10442	 & 0.10613	 & 0.10523	 & 73.253	 & 0.067	 & 0.30	 \\
		0.10613	 & 0.10783	 & 0.10693	 & 70.749	 & 0.064	 & 0.28	 \\
		0.10783	 & 0.10954	 & 0.10864	 & 68.344	 & 0.062	 & 0.27	 \\
		0.10954	 & 0.11125	 & 0.11035	 & 66.073	 & 0.060	 & 0.26	 \\
		0.11125	 & 0.11296	 & 0.11206	 & 63.812	 & 0.058	 & 0.24	 \\
		0.11296	 & 0.11468	 & 0.11377	 & 61.604	 & 0.056	 & 0.23	 \\
		0.11468	 & 0.11639	 & 0.11549	 & 59.525	 & 0.054	 & 0.22	 \\
		0.11639	 & 0.11811	 & 0.11720	 & 57.524	 & 0.052	 & 0.21	 \\
		0.11811	 & 0.11983	 & 0.11892	 & 55.529	 & 0.050	 & 0.20	 \\
		0.11983	 & 0.12154	 & 0.12064	 & 53.648	 & 0.049	 & 0.19	 \\
		0.12154	 & 0.12327	 & 0.12236	 & 51.755	 & 0.047	 & 0.18	 \\
		0.12327	 & 0.12499	 & 0.12408	 & 50.026	 & 0.045	 & 0.17	 \\
		0.12499	 & 0.12671	 & 0.12580	 & 48.342	 & 0.044	 & 0.16	 \\
		0.12671	 & 0.12844	 & 0.12753	 & 46.655	 & 0.042	 & 0.16	 \\
\hline 
			            \end{tabular}
        \label{analysis_corrections}
        \end{center}
    \end{table*}

    \begin{table*}\small\color{black}
        \begin{center}
            \caption{Continuation of Table 5.}
            \begin{tabular}{ | c  c  c  c  c  c|}
		\hline
			$|t|_{\rm low}$		& $|t|_{\rm high}$		&	$|t|_{\rm repr.}$		&	${\rm d}\sigma/{\rm d}t$	& Statistical uncertainty 	&	Systematic uncertainty \\ 
						& [GeV$^{2}$]			&					&					&	[mb GeV$^{-2}$]		& \\ \hline 

		0.12844	 & 0.13017	 & 0.12926	 & 45.0718	 & 0.0409	 & 0.148	 \\
		0.13017	 & 0.13190	 & 0.13099	 & 43.5530	 & 0.0396	 & 0.141	 \\
		0.13190	 & 0.13363	 & 0.13272	 & 42.0412	 & 0.0383	 & 0.134	 \\
		0.13363	 & 0.13537	 & 0.13445	 & 40.6083	 & 0.0370	 & 0.127	 \\
		0.13537	 & 0.13710	 & 0.13618	 & 39.1987	 & 0.0357	 & 0.120	 \\
		0.13710	 & 0.13884	 & 0.13792	 & 37.8674	 & 0.0346	 & 0.114	 \\
		0.13884	 & 0.14058	 & 0.13966	 & 36.5563	 & 0.0334	 & 0.108	 \\
		0.14058	 & 0.14232	 & 0.14140	 & 35.3093	 & 0.0323	 & 0.104	 \\
		0.14232	 & 0.14406	 & 0.14314	 & 34.0693	 & 0.0312	 & 0.099	 \\
		0.14406	 & 0.14580	 & 0.14488	 & 32.8900	 & 0.0302	 & 0.095	 \\
		0.14580	 & 0.14755	 & 0.14663	 & 31.7420	 & 0.0292	 & 0.091	 \\
		0.14755	 & 0.14930	 & 0.14837	 & 30.6370	 & 0.0282	 & 0.088	 \\
		0.14930	 & 0.15105	 & 0.15012	 & 29.5747	 & 0.0273	 & 0.084	 \\
		0.15105	 & 0.15280	 & 0.15187	 & 28.5268	 & 0.0264	 & 0.080	 \\
		0.15280	 & 0.15455	 & 0.15363	 & 27.5185	 & 0.0256	 & 0.077	 \\
		0.15455	 & 0.15630	 & 0.15538	 & 26.5863	 & 0.0247	 & 0.074	 \\
		0.15630	 & 0.15806	 & 0.15713	 & 25.6414	 & 0.0239	 & 0.071	 \\
		0.15806	 & 0.15982	 & 0.15889	 & 24.7246	 & 0.0231	 & 0.068	 \\
		0.15982	 & 0.16158	 & 0.16065	 & 23.8487	 & 0.0224	 & 0.065	 \\
		0.16158	 & 0.16334	 & 0.16241	 & 22.9954	 & 0.0216	 & 0.062	 \\
		0.16334	 & 0.16510	 & 0.16417	 & 22.1702	 & 0.0209	 & 0.059	 \\
		0.16510	 & 0.16687	 & 0.16594	 & 21.4001	 & 0.0203	 & 0.057	 \\
		0.16687	 & 0.16864	 & 0.16770	 & 20.6409	 & 0.0196	 & 0.054	 \\
		0.16864	 & 0.17040	 & 0.16947	 & 19.9019	 & 0.0190	 & 0.052	 \\
		0.17040	 & 0.17218	 & 0.17124	 & 19.1646	 & 0.0183	 & 0.050	 \\
		0.17218	 & 0.17395	 & 0.17301	 & 18.4862	 & 0.0177	 & 0.048	 \\
		0.17395	 & 0.17572	 & 0.17478	 & 17.7879	 & 0.0172	 & 0.045	 \\
		0.17572	 & 0.17750	 & 0.17656	 & 17.1552	 & 0.0166	 & 0.043	 \\
		0.17750	 & 0.17927	 & 0.17834	 & 16.5560	 & 0.0161	 & 0.042	 \\
		0.17927	 & 0.18105	 & 0.18011	 & 15.9388	 & 0.0155	 & 0.040	 \\
		0.18105	 & 0.18283	 & 0.18189	 & 15.3767	 & 0.0151	 & 0.038	 \\
		0.18283	 & 0.18462	 & 0.18368	 & 14.8025	 & 0.0146	 & 0.036	 \\
		0.18462	 & 0.18640	 & 0.18546	 & 14.2666	 & 0.0141	 & 0.035	 \\
		0.18640	 & 0.18819	 & 0.18725	 & 13.7533	 & 0.0137	 & 0.033	 \\
		0.18819	 & 0.18998	 & 0.18903	 & 13.2581	 & 0.0132	 & 0.032	 \\
		0.18998	 & 0.19177	 & 0.19082	 & 12.7590	 & 0.0128	 & 0.030	 \\
		0.19177	 & 0.19356	 & 0.19261	 & 12.2949	 & 0.0124	 & 0.029	 \\
		0.19356	 & 0.19535	 & 0.19440	 & 11.8561	 & 0.0120	 & 0.028	 \\
		0.19535	 & 0.19715	 & 0.19620	 & 11.4076	 & 0.0116	 & 0.026	 \\
		0.19715	 & 0.19894	 & 0.19800	 & 10.9936	 & 0.0113	 & 0.025	 \\
		0.19894	 & 0.20074	 & 0.19979	 & 10.5950	 & 0.0109	 & 0.024	 \\
		0.20074	 & 0.20254	 & 0.20159	 & 10.1833	 & 0.0106	 & 0.023	 \\
		0.20254	 & 0.20435	 & 0.20339	 & 9.8035	 & 0.0102	 & 0.022	 \\
		0.20435	 & 0.20615	 & 0.20520	 & 9.4488	 & 0.0099	 & 0.021	 \\
		0.20615	 & 0.20796	 & 0.20700	 & 9.1026	 & 0.0096	 & 0.020	 \\
		0.20796	 & 0.20977	 & 0.20881	 & 8.7620	 & 0.0093	 & 0.019	 \\
		0.20977	 & 0.21158	 & 0.21062	 & 8.4083	 & 0.0090	 & 0.018	 \\
		0.21158	 & 0.21339	 & 0.21243	 & 8.1080	 & 0.0087	 & 0.017	 \\
		0.21339	 & 0.21520	 & 0.21424	 & 7.8116	 & 0.0084	 & 0.016	 \\
		0.21520	 & 0.21702	 & 0.21606	 & 7.5264	 & 0.0082	 & 0.016	 \\
		0.21702	 & 0.21883	 & 0.21787	 & 7.2372	 & 0.0079	 & 0.015	 \\
		0.21883	 & 0.22065	 & 0.21969	 & 6.9655	 & 0.0077	 & 0.014	 \\
		0.22065	 & 0.22247	 & 0.22151	 & 6.7154	 & 0.0075	 & 0.013	 \\
		0.22247	 & 0.22429	 & 0.22333	 & 6.4637	 & 0.0072	 & 0.013	 \\
\hline 
			            \end{tabular}
        \label{analysis_corrections}
        \end{center}
    \end{table*}

    \begin{table*}\small\color{black}
        \begin{center}
            \caption{Continuation of Table 5.}
            \begin{tabular}{ | c  c  c  c  c  c|}
		\hline
			$|t|_{\rm low}$		& $|t|_{\rm high}$		&	$|t|_{\rm repr.}$		&	${\rm d}\sigma/{\rm d}t$	& Statistical uncertainty 	&	Systematic uncertainty \\ 
						& [GeV$^{2}$]			&					&					&	[mb GeV$^{-2}$]		& \\ \hline 

		0.22429	 & 0.22612	 & 0.22515	 & 6.20908	 & 0.0070	 & 0.0122	 \\
		0.22612	 & 0.22794	 & 0.22698	 & 5.97031	 & 0.0068	 & 0.0116	 \\
		0.22794	 & 0.22977	 & 0.22881	 & 5.75469	 & 0.0066	 & 0.0111	 \\
		0.22977	 & 0.23160	 & 0.23064	 & 5.53763	 & 0.0064	 & 0.0105	 \\
		0.23160	 & 0.23343	 & 0.23247	 & 5.32432	 & 0.0062	 & 0.0101	 \\
		0.23343	 & 0.23527	 & 0.23430	 & 5.11779	 & 0.0060	 & 0.0097	 \\
		0.23527	 & 0.23710	 & 0.23613	 & 4.91989	 & 0.0058	 & 0.0093	 \\
		0.23710	 & 0.23894	 & 0.23797	 & 4.72926	 & 0.0057	 & 0.0089	 \\
		0.23894	 & 0.24078	 & 0.23981	 & 4.54239	 & 0.0055	 & 0.0085	 \\
		0.24078	 & 0.24262	 & 0.24165	 & 4.37635	 & 0.0053	 & 0.0081	 \\
		0.24262	 & 0.24446	 & 0.24349	 & 4.20548	 & 0.0052	 & 0.0078	 \\
		0.24446	 & 0.24631	 & 0.24533	 & 4.03536	 & 0.0050	 & 0.0075	 \\
		0.24631	 & 0.24815	 & 0.24718	 & 3.89381	 & 0.0049	 & 0.0071	 \\
		0.24815	 & 0.25000	 & 0.24902	 & 3.73005	 & 0.0047	 & 0.0068	 \\
		0.25000	 & 0.25185	 & 0.25087	 & 3.58604	 & 0.0046	 & 0.0065	 \\
		0.25185	 & 0.25370	 & 0.25272	 & 3.44067	 & 0.0044	 & 0.0062	 \\
		0.25370	 & 0.25556	 & 0.25458	 & 3.30493	 & 0.0043	 & 0.0060	 \\
		0.25556	 & 0.25741	 & 0.25643	 & 3.17651	 & 0.0042	 & 0.0057	 \\
		0.25741	 & 0.25927	 & 0.25829	 & 3.04288	 & 0.0040	 & 0.0055	 \\
		0.25927	 & 0.26113	 & 0.26015	 & 2.92947	 & 0.0039	 & 0.0052	 \\
		0.26113	 & 0.26299	 & 0.26201	 & 2.80257	 & 0.0038	 & 0.0050	 \\
		0.26299	 & 0.26485	 & 0.26387	 & 2.69417	 & 0.0037	 & 0.0048	 \\
		0.26485	 & 0.26672	 & 0.26573	 & 2.57945	 & 0.0036	 & 0.0046	 \\
		0.26672	 & 0.26858	 & 0.26760	 & 2.47640	 & 0.0035	 & 0.0044	 \\
		0.26858	 & 0.27045	 & 0.26947	 & 2.37881	 & 0.0034	 & 0.0042	 \\
		0.27045	 & 0.27232	 & 0.27134	 & 2.27509	 & 0.0033	 & 0.0040	 \\
		0.27232	 & 0.27420	 & 0.27321	 & 2.18539	 & 0.0032	 & 0.0038	 \\
		0.27420	 & 0.27607	 & 0.27508	 & 2.09632	 & 0.0031	 & 0.0036	 \\
		0.27607	 & 0.27795	 & 0.27695	 & 2.00623	 & 0.0030	 & 0.0035	 \\
		0.27795	 & 0.27982	 & 0.27883	 & 1.92209	 & 0.0029	 & 0.0033	 \\
		0.27982	 & 0.28170	 & 0.28071	 & 1.84416	 & 0.0028	 & 0.0031	 \\
		0.28170	 & 0.28359	 & 0.28259	 & 1.76460	 & 0.0027	 & 0.0030	 \\
		0.28359	 & 0.28547	 & 0.28447	 & 1.69359	 & 0.0027	 & 0.0029	 \\
		0.28547	 & 0.28735	 & 0.28636	 & 1.62036	 & 0.0026	 & 0.0027	 \\
		0.28735	 & 0.28924	 & 0.28824	 & 1.54873	 & 0.0025	 & 0.0026	 \\
		0.28924	 & 0.29113	 & 0.29013	 & 1.48620	 & 0.0024	 & 0.0025	 \\
		0.29113	 & 0.29302	 & 0.29202	 & 1.42084	 & 0.0024	 & 0.0024	 \\
		0.29302	 & 0.29492	 & 0.29391	 & 1.35815	 & 0.0023	 & 0.0022	 \\
		0.29492	 & 0.29681	 & 0.29581	 & 1.29937	 & 0.0022	 & 0.0021	 \\
		0.29681	 & 0.29871	 & 0.29770	 & 1.24334	 & 0.0022	 & 0.0020	 \\
		0.29871	 & 0.30061	 & 0.29960	 & 1.18712	 & 0.0021	 & 0.0019	 \\
		0.30061	 & 0.30251	 & 0.30150	 & 1.13726	 & 0.0021	 & 0.0018	 \\
		0.30251	 & 0.30441	 & 0.30340	 & 1.08606	 & 0.0020	 & 0.0018	 \\
		0.30441	 & 0.30631	 & 0.30531	 & 1.03785	 & 0.0019	 & 0.0017	 \\
		0.30631	 & 0.30822	 & 0.30721	 & 0.99128	 & 0.0019	 & 0.0016	 \\
		0.30822	 & 0.31013	 & 0.30912	 & 0.94404	 & 0.0018	 & 0.0015	 \\
		0.31013	 & 0.31204	 & 0.31103	 & 0.90267	 & 0.0018	 & 0.0014	 \\
		0.31204	 & 0.31395	 & 0.31294	 & 0.86103	 & 0.0017	 & 0.0014	 \\
		0.31395	 & 0.31586	 & 0.31485	 & 0.82014	 & 0.0017	 & 0.0013	 \\
		0.31586	 & 0.31778	 & 0.31677	 & 0.78056	 & 0.0016	 & 0.0012	 \\
		0.31778	 & 0.31970	 & 0.31868	 & 0.74955	 & 0.0016	 & 0.0012	 \\
		0.31970	 & 0.32162	 & 0.32060	 & 0.71498	 & 0.0015	 & 0.0011	 \\
		0.32162	 & 0.32354	 & 0.32252	 & 0.67820	 & 0.0015	 & 0.0010	 \\
		0.32354	 & 0.32546	 & 0.32445	 & 0.64830	 & 0.0014	 & 0.0010	 \\
\hline 
			            \end{tabular}
        \label{analysis_corrections}
        \end{center}
    \end{table*}

    \begin{table*}\small\color{black}
        \begin{center}
            \caption{Continuation of Table 5.}
            \begin{tabular}{ | c  c  c  c  c  c|}
		\hline
			$|t|_{\rm low}$		& $|t|_{\rm high}$		&	$|t|_{\rm repr.}$		&	${\rm d}\sigma/{\rm d}t$	& Statistical uncertainty 	&	Systematic uncertainty \\ 
						& [GeV$^{2}$]			&					&					&	[mb GeV$^{-2}$]		& \\ \hline 

		0.32546	 & 0.32739	 & 0.32637	 & 0.6160172	 & 0.00140	 & 0.00094	 \\
		0.32739	 & 0.32932	 & 0.32830	 & 0.5875874	 & 0.00136	 & 0.00090	 \\
		0.32932	 & 0.33124	 & 0.33022	 & 0.5599290	 & 0.00132	 & 0.00086	 \\
		0.33124	 & 0.33318	 & 0.33215	 & 0.5326184	 & 0.00129	 & 0.00081	 \\
		0.33318	 & 0.33511	 & 0.33409	 & 0.5084319	 & 0.00125	 & 0.00077	 \\
		0.33511	 & 0.33704	 & 0.33602	 & 0.4841120	 & 0.00121	 & 0.00074	 \\
		0.33704	 & 0.33898	 & 0.33796	 & 0.4584498	 & 0.00118	 & 0.00070	 \\
		0.33898	 & 0.34092	 & 0.33989	 & 0.4368558	 & 0.00114	 & 0.00067	 \\
		0.34092	 & 0.34286	 & 0.34183	 & 0.4159908	 & 0.00111	 & 0.00063	 \\
		0.34286	 & 0.34480	 & 0.34378	 & 0.3931062	 & 0.00108	 & 0.00060	 \\
		0.34480	 & 0.34675	 & 0.34572	 & 0.3773297	 & 0.00105	 & 0.00057	 \\
		0.34675	 & 0.34870	 & 0.34767	 & 0.3567071	 & 0.00102	 & 0.00054	 \\
		0.34870	 & 0.35064	 & 0.34961	 & 0.3408495	 & 0.00099	 & 0.00052	 \\
		0.35064	 & 0.35259	 & 0.35156	 & 0.3233294	 & 0.00096	 & 0.00049	 \\
		0.35259	 & 0.35455	 & 0.35351	 & 0.3078857	 & 0.00093	 & 0.00047	 \\
		0.35455	 & 0.35650	 & 0.35547	 & 0.2911156	 & 0.00090	 & 0.00044	 \\
		0.35650	 & 0.35846	 & 0.35742	 & 0.2759747	 & 0.00088	 & 0.00042	 \\
		0.35846	 & 0.36042	 & 0.35938	 & 0.2632198	 & 0.00085	 & 0.00040	 \\
		0.36042	 & 0.36238	 & 0.36134	 & 0.2484736	 & 0.00083	 & 0.00038	 \\
		0.36238	 & 0.36434	 & 0.36330	 & 0.2368378	 & 0.00080	 & 0.00036	 \\
		0.36434	 & 0.36630	 & 0.36526	 & 0.2238501	 & 0.00078	 & 0.00035	 \\
		0.36630	 & 0.36827	 & 0.36723	 & 0.2129791	 & 0.00076	 & 0.00034	 \\
		0.36827	 & 0.37024	 & 0.36920	 & 0.2001638	 & 0.00073	 & 0.00032	 \\
		0.37024	 & 0.37221	 & 0.37117	 & 0.1910685	 & 0.00071	 & 0.00031	 \\
		0.37221	 & 0.37418	 & 0.37314	 & 0.1811450	 & 0.00069	 & 0.00030	 \\
		0.37418	 & 0.37616	 & 0.37511	 & 0.1708701	 & 0.00067	 & 0.00029	 \\
		0.37616	 & 0.37813	 & 0.37709	 & 0.1631546	 & 0.00065	 & 0.00028	 \\
		0.37813	 & 0.38011	 & 0.37906	 & 0.1542701	 & 0.00063	 & 0.00027	 \\
		0.38011	 & 0.38209	 & 0.38104	 & 0.1458654	 & 0.00061	 & 0.00025	 \\
		0.38209	 & 0.38407	 & 0.38302	 & 0.1376252	 & 0.00059	 & 0.00024	 \\
		0.38407	 & 0.38606	 & 0.38501	 & 0.1314138	 & 0.00058	 & 0.00023	 \\
		0.38606	 & 0.38804	 & 0.38699	 & 0.1250128	 & 0.00056	 & 0.00022	 \\
		0.38804	 & 0.39003	 & 0.38898	 & 0.1172248	 & 0.00054	 & 0.00020	 \\
		0.39003	 & 0.39202	 & 0.39097	 & 0.1108358	 & 0.00053	 & 0.00019	 \\
		0.39202	 & 0.39401	 & 0.39296	 & 0.1051992	 & 0.00051	 & 0.00018	 \\
		0.39401	 & 0.39601	 & 0.39495	 & 0.0988553	 & 0.00049	 & 0.00017	 \\
		0.39601	 & 0.39800	 & 0.39695	 & 0.0943577	 & 0.00048	 & 0.00016	 \\
		0.39800	 & 0.40000	 & 0.39894	 & 0.0893882	 & 0.00047	 & 0.00014	 \\
		0.40000	 & 0.40200	 & 0.40094	 & 0.0844118	 & 0.00045	 & 0.00013	 \\
		0.40200	 & 0.40410	 & 0.40299	 & 0.0794673	 & 0.00043	 & 0.00012	 \\
		0.40410	 & 0.40631	 & 0.40514	 & 0.0746659	 & 0.00040	 & 0.00011	 \\
		0.40631	 & 0.40866	 & 0.40742	 & 0.0702312	 & 0.00038	 & 0.00010	 \\
		0.40866	 & 0.41117	 & 0.40984	 & 0.0644914	 & 0.00035	 & 0.00008	 \\
		0.41117	 & 0.41384	 & 0.41242	 & 0.0612570	 & 0.00033	 & 0.00008	 \\
		0.41384	 & 0.41668	 & 0.41517	 & 0.0562343	 & 0.00030	 & 0.00008	 \\
		0.41668	 & 0.41974	 & 0.41812	 & 0.0526126	 & 0.00028	 & 0.00007	 \\
		0.41974	 & 0.42305	 & 0.42129	 & 0.0483678	 & 0.00026	 & 0.00007	 \\
		0.42305	 & 0.42663	 & 0.42473	 & 0.0449274	 & 0.00024	 & 0.00011	 \\
		0.42663	 & 0.43054	 & 0.42846	 & 0.0413754	 & 0.00022	 & 0.00022	 \\
		0.43054	 & 0.43483	 & 0.43255	 & 0.0385490	 & 0.00020	 & 0.00038	 \\
		0.43483	 & 0.43957	 & 0.43705	 & 0.0354994	 & 0.00019	 & 0.00061	 \\
		0.43957	 & 0.44485	 & 0.44204	 & 0.0327863	 & 0.00017	 & 0.00093	 \\
		0.44485	 & 0.45072	 & 0.44760	 & 0.0307740	 & 0.00016	 & 0.00122	 \\
		0.45072	 & 0.45722	 & 0.45377	 & 0.0291248	 & 0.00015	 & 0.00084	 \\
\hline 
			            \end{tabular}
        \label{analysis_corrections}
        \end{center}
    \end{table*}

    \begin{table*}\small\color{black}
        \begin{center}
            \caption{Continuation of Table 5.}
            \begin{tabular}{ | c  c  c  c  c  c|}
		\hline
			$|t|_{\rm low}$		& $|t|_{\rm high}$		&	$|t|_{\rm repr.}$		&	${\rm d}\sigma/{\rm d}t$	& Statistical uncertainty 	&	Systematic uncertainty \\ 
						& [GeV$^{2}$]			&					&					&	[mb GeV$^{-2}$]		& \\ \hline 

		0.45722	 & 0.46437	 & 0.46059	 & 0.0277333	 & 0.00015	 & 0.00027	 \\
		0.46437	 & 0.47203	 & 0.46799	 & 0.0272682	 & 0.00014	 & 0.00032	 \\
		0.47203	 & 0.48006	 & 0.47585	 & 0.0276871	 & 0.00015	 & 0.00035	 \\
		0.48006	 & 0.48828	 & 0.48398	 & 0.0284679	 & 0.00015	 & 0.00042	 \\
		0.48828	 & 0.49654	 & 0.49222	 & 0.0294984	 & 0.00015	 & 0.00051	 \\
		0.49654	 & 0.50472	 & 0.50045	 & 0.0305493	 & 0.00016	 & 0.00054	 \\
		0.50472	 & 0.51278	 & 0.50857	 & 0.0320606	 & 0.00017	 & 0.00053	 \\
		0.51278	 & 0.52069	 & 0.51656	 & 0.0337122	 & 0.00018	 & 0.00045	 \\
		0.52069	 & 0.52845	 & 0.52440	 & 0.0352345	 & 0.00019	 & 0.00034	 \\
		0.52845	 & 0.53607	 & 0.53209	 & 0.0368763	 & 0.00020	 & 0.00024	 \\
		0.53607	 & 0.54356	 & 0.53964	 & 0.0382915	 & 0.00020	 & 0.00018	 \\
		0.54356	 & 0.55093	 & 0.54707	 & 0.0396758	 & 0.00021	 & 0.00019	 \\
		0.55093	 & 0.55820	 & 0.55440	 & 0.0411312	 & 0.00022	 & 0.00019	 \\
		0.55820	 & 0.56539	 & 0.56163	 & 0.0427254	 & 0.00022	 & 0.00021	 \\
		0.56539	 & 0.57250	 & 0.56878	 & 0.0434895	 & 0.00023	 & 0.00022	 \\
		0.57250	 & 0.57956	 & 0.57587	 & 0.0444147	 & 0.00023	 & 0.00024	 \\
		0.57956	 & 0.58657	 & 0.58290	 & 0.0454339	 & 0.00024	 & 0.00022	 \\
		0.58657	 & 0.59356	 & 0.58990	 & 0.0462303	 & 0.00024	 & 0.00020	 \\
		0.59356	 & 0.60052	 & 0.59687	 & 0.0462224	 & 0.00024	 & 0.00019	 \\
		0.60052	 & 0.60748	 & 0.60383	 & 0.0468683	 & 0.00025	 & 0.00018	 \\
		0.60748	 & 0.61444	 & 0.61079	 & 0.0471657	 & 0.00025	 & 0.00017	 \\
		0.61444	 & 0.62142	 & 0.61776	 & 0.0479109	 & 0.00025	 & 0.00017	 \\
		0.62142	 & 0.62844	 & 0.62476	 & 0.0477659	 & 0.00025	 & 0.00016	 \\
		0.62844	 & 0.63550	 & 0.63179	 & 0.0483139	 & 0.00025	 & 0.00016	 \\
		0.63550	 & 0.64262	 & 0.63888	 & 0.0479224	 & 0.00025	 & 0.00016	 \\
		0.64262	 & 0.64981	 & 0.64603	 & 0.0478640	 & 0.00025	 & 0.00015	 \\
		0.64981	 & 0.65710	 & 0.65327	 & 0.0479463	 & 0.00025	 & 0.00015	 \\
		0.65710	 & 0.66449	 & 0.66060	 & 0.0473400	 & 0.00025	 & 0.00015	 \\
		0.66449	 & 0.67200	 & 0.66805	 & 0.0475291	 & 0.00025	 & 0.00014	 \\
		0.67200	 & 0.67966	 & 0.67563	 & 0.0466393	 & 0.00025	 & 0.00014	 \\
		0.67966	 & 0.68747	 & 0.68336	 & 0.0460401	 & 0.00024	 & 0.00014	 \\
		0.68747	 & 0.69545	 & 0.69125	 & 0.0452852	 & 0.00024	 & 0.00014	 \\
		0.69545	 & 0.70363	 & 0.69933	 & 0.0446399	 & 0.00024	 & 0.00013	 \\
		0.70363	 & 0.71203	 & 0.70761	 & 0.0440867	 & 0.00023	 & 0.00013	 \\
		0.71203	 & 0.72066	 & 0.71611	 & 0.0429611	 & 0.00023	 & 0.00013	 \\
		0.72066	 & 0.72956	 & 0.72487	 & 0.0419997	 & 0.00022	 & 0.00012	 \\
		0.72956	 & 0.73875	 & 0.73391	 & 0.0415542	 & 0.00022	 & 0.00012	 \\
		0.73875	 & 0.74826	 & 0.74324	 & 0.0408855	 & 0.00021	 & 0.00012	 \\
		0.74826	 & 0.75812	 & 0.75292	 & 0.0394498	 & 0.00021	 & 0.00011	 \\
		0.75812	 & 0.76837	 & 0.76296	 & 0.0384823	 & 0.00020	 & 0.00011	 \\
		0.76837	 & 0.77905	 & 0.77341	 & 0.0368327	 & 0.00019	 & 0.00011	 \\
		0.77905	 & 0.79021	 & 0.78432	 & 0.0357022	 & 0.00019	 & 0.00010	 \\
		0.79021	 & 0.80190	 & 0.79572	 & 0.0344881	 & 0.00018	 & 0.00010	 \\
		0.80190	 & 0.81419	 & 0.80769	 & 0.0331379	 & 0.00017	 & 0.00009	 \\
		0.81419	 & 0.82715	 & 0.82029	 & 0.0317622	 & 0.00017	 & 0.00009	 \\
		0.82715	 & 0.84087	 & 0.83361	 & 0.0301535	 & 0.00016	 & 0.00008	 \\
		0.84087	 & 0.85546	 & 0.84773	 & 0.0286175	 & 0.00015	 & 0.00008	 \\
		0.85546	 & 0.87105	 & 0.86279	 & 0.0268686	 & 0.00014	 & 0.00007	 \\
		0.87105	 & 0.88781	 & 0.87892	 & 0.0257198	 & 0.00013	 & 0.00007	 \\
		0.88781	 & 0.90598	 & 0.89633	 & 0.0236250	 & 0.00012	 & 0.00006	 \\
		0.90598	 & 0.92585	 & 0.91528	 & 0.0217698	 & 0.00012	 & 0.00006	 \\
		0.92585	 & 0.94787	 & 0.93614	 & 0.0199802	 & 0.00011	 & 0.00005	 \\
		0.94787	 & 0.97267	 & 0.95942	 & 0.0179976	 & 0.00010	 & 0.00005	 \\
		0.97267	 & 1.00119	 & 0.98591	 & 0.0159908	 & 0.00008	 & 0.00004	 \\
\hline 
			            \end{tabular}
        \label{analysis_corrections}
        \end{center}
    \end{table*}

    \begin{table*}\small\color{black}
        \begin{center}
            \caption{Continuation of Table 5.}
            \begin{tabular}{ | c  c  c  c  c  c|}
		\hline
			$|t|_{\rm low}$		& $|t|_{\rm high}$		&	$|t|_{\rm repr.}$		&	${\rm d}\sigma/{\rm d}t$	& Statistical uncertainty 	&	Systematic uncertainty \\ 
						& [GeV$^{2}$]			&					&					&	[mb GeV$^{-2}$]		& \\ \hline 

		1.00119	 & 1.03483	 & 1.01673	 & 0.01377598	 & 0.00007261	 & 0.000036737	 \\
		1.03483	 & 1.07580	 & 1.05364	 & 0.01160554	 & 0.00006101	 & 0.000030768	 \\
		1.07580	 & 1.12787	 & 1.09948	 & 0.00933244	 & 0.00004911	 & 0.000024537	 \\
		1.12787	 & 1.19956	 & 1.15991	 & 0.00700284	 & 0.00003679	 & 0.000018286	 \\
		1.19956	 & 1.27842	 & 1.23455	 & 0.00478260	 & 0.00002953	 & 0.000012760	 \\
		1.27842	 & 1.36517	 & 1.31661	 & 0.00323285	 & 0.00002357	 & 0.000008191	 \\
		1.36517	 & 1.46060	 & 1.40681	 & 0.00206329	 & 0.00001825	 & 0.000005511	 \\
		1.46060	 & 1.56556	 & 1.50598	 & 0.00125878	 & 0.00001387	 & 0.000003100	 \\
		1.56556	 & 1.68102	 & 1.61503	 & 0.00072459	 & 0.00001022	 & 0.000002075	 \\
		1.68102	 & 1.80803	 & 1.73499	 & 0.00040267	 & 0.00000740	 & 0.000001886	 \\
		1.80803	 & 1.94774	 & 1.86706	 & 0.00021747	 & 0.00000530	 & 0.000000856	 \\
		1.94774	 & 2.10142	 & 2.01215	 & 0.00010673	 & 0.00000360	 & 0.000000473	 \\
		2.10142	 & 2.27047	 & 2.17127	 & 0.00005870	 & 0.00000260	 & 0.000000276	 \\
		2.27047	 & 2.45642	 & 2.34561	 & 0.00002434	 & 0.00000163	 & 0.000000114	 \\
		2.45642	 & 2.66097	 & 2.53863	 & 0.00001017	 & 0.00000102	 & 0.000000215	 \\
		2.66097	 & 2.88597	 & 2.75357	 & 0.00000395	 & 0.00000063	 & 0.000000055	 \\
		2.88597	 & 3.13348	 & 2.98393	 & 0.00000235	 & 0.00000050	 & 0.000000031	 \\
		3.13348	 & 3.40573	 & 3.23982	 & 0.00000051	 & 0.00000023	 & 0.000000003	 \\
		3.40573	 & 3.70521	 & 3.52028	 & 0.00000029	 & 0.00000017	 & 0.000000002	 \\
		3.70521	 & 4.03464	 & 3.82873	 & 0.00000020	 & 0.00000012	 & 0.000000003	 \\\hline
			            \end{tabular}
        \label{analysis_corrections}
        \end{center}
    \end{table*}

\section*{Acknowledgments}

We are grateful to the beam optics development team
for the design and the successful commissioning of the
high $\beta^{*}$ optics and to the LHC machine coordinators for
scheduling the dedicated fills.

This work was supported by the institutions listed on the
front page and partially also by NSF (US), the Magnus Ehrnrooth Foundation (Finland), the Waldemar von
Frenckell Foundation (Finland), the Academy of Finland,
the Finnish Academy of Science and Letters (The Vilho 
Yrj\"o and Kalle V\"ais\"al\"a Fund), the Circles of Knowledge Club (Hungary) and the OTKA NK 101438 and the EFOP-3.6.1-16-2016-00001 grants
(Hungary). Individuals have received support from Nylands nation vid Helsingfors universitet (Finland),
MSMT CR (the Czech Republic), the J\'{a}nos Bolyai Research Scholarship of
the Hungarian Academy of Sciences, the NKP-17-4 New National Excellence Program of the
Hungarian Ministry of Human Capacities and the Polish Ministry of Science and Higher Education
Grant no. MNiSW DIR/WK/2017/07-01.

\clearpage
\bibliographystyle{utphys}
\bibliography{mybib}

\end{document}